\documentclass[aps,nofootinbib,superscriptaddress]{revtex4-1}
 \pdfoutput=1
\usepackage[a4paper,left=1.5cm,right=1.5cm,top=3cm,bottom=3cm]{geometry}
\usepackage[toc]{appendix} 

\usepackage{amssymb,amsmath,amsfonts}
\usepackage{xfrac}
\usepackage[dvipsnames]{xcolor}
\usepackage{graphicx}
\usepackage{longtable}
\usepackage{verbatim}
\usepackage{color}
\usepackage{mdframed}
\usepackage{color}
\usepackage{framed}
\usepackage{hyperref}
\usepackage{multirow}
\hypersetup{colorlinks, citecolor=bluscuro, linkcolor=black, urlcolor=bluscuro}
\definecolor{rossos}{cmyk}{0,1,1,0.55}
\definecolor{bluscuro}{rgb}{0.15, 0.2, .85}
\definecolor{bluchiaro}{cmyk}{1,.3,0.,0.1}

\graphicspath{{./Figures/}}

\newcommand{\be}{\begin{equation}}
\newcommand{\ee}{\end{equation}}
\newcommand{\bea}{\begin{eqnarray}}
\newcommand{\eea}{\end{eqnarray}}
\newcommand{\beq}{\begin{equation}}
\newcommand{\eeq}{\end{equation}}
\def\beqa{\begin{eqnarray}}

\def\vx{{\vec{x}}}
\def\vk{{\vec{k}}}
\def\vy{{\vec{y}}}

\def\e{\delta\rho}
\def\d{{\rm d}}
\def\hr{{\hat r}}
\def\NL{\text{\tiny NL}}
\def\G{\text{\tiny G}}
\def\NG{\text{\tiny NG}}
\def\PBH{\text{\tiny PBH}}
\def\eeqa{\end{eqnarray}}

\def\lsim{\mathrel{\rlap{\lower4pt\hbox{\hskip0.5pt$\sim$}}
    \raise1pt\hbox{$<$}}}         
\def\gsim{\mathrel{\rlap{\lower4pt\hbox{\hskip0.5pt$\sim$}}
    \raise1pt\hbox{$>$}}}         

\usepackage[normalem]{ulem}

\newcommand{\arXiv}[2]{\href{http://arxiv.org/pdf/#1}{{\tt [#2/#1]}}}
\newcommand{\arXivold}[1]{\href{http://arxiv.org/pdf/#1}{{\tt [#1]}}}
\numberwithin{equation}{section}

\begin{document}
\vskip 5cm

\title{Non-Gaussian Formation of Primordial Black Holes:  Effects on the Threshold }

\author{A. Kehagias }
\address{Physics Division, National Technical University of Athens, 15780 Zografou Campus, Athens, Greece}
\email{kehagias@central.ntua.gr}

\author{I. Musco}
\address{Institut de Ci\`encies del Cosmos, Universitat de Barcelona, 
Mart\'i i Franqu\`es 1, 08028 Barcelona, Spain}
\email{iliamusco@icc.ub.edu}

\author{A.~Riotto}
\address{D\'epartement de Physique Th\'eorique and Centre for Astroparticle Physics (CAP), Universit\'e de Gen\`eve, 24 quai 
E. Ansermet, CH-1211 Geneva, Switzerland}
\email{antonio.riotto@unige.ch}
\address{CERN,
Theoretical Physics Department, Geneva, Switzerland}


\begin{abstract}
\noindent
Primordial black holes could have been formed in the early universe from sufficiently large cosmological perturbations re-entering 
the horizon when the Universe is still radiation dominated. These originate from the spectrum of curvature perturbations generated 
during inflation at small-scales. Because of the non-linear relation between the curvature perturbation $\zeta$ and the overdensity 
$\delta\rho$, the formation of the primordial black holes is affected by intrinsic non-Gaussianity even if the curvature perturbation is 
Gaussian. We investigate the impact of this non-Gaussianity on the critical threshold $\delta_c$ which measures the excess of mass 
of the perturbation, finding a relative change with respect to the value obtained using a linear relation between $\zeta$ and $\delta\rho$, 
of a few percent suggesting that the value of the critical threshold is rather robust against non-linearities. The same holds also when local 
primordial non-Gaussianity, with $f_\NL\gsim-3/2$, are added to the curvature perturbation.
\end{abstract}

\maketitle

\tableofcontents

\section{Introduction}
\label{Introduction}
\noindent
Primordial Black Holes (PBHs) have recently received much attention starting from the discovery of the gravitational waves emitted by 
the merging  of two $\sim 30\, M_\odot$ black holes \cite{ligo1}. In particular, the focus has been on the possibility that PBHs may 
describe the nature of dark matter we observe in the universe \cite{bird} (see also \cite{revPBH} and references therein). 

Even though there are many ways to generate PBHs in the early universe, the mechanism which has been investigated more extensively
in the recent literature is the one obtained from inflation \cite{s1,s2,s3}. During such stage of primordial acceleration, the curvature
perturbation $\zeta$ may be enhanced at small-scales with respect to the large-scale perturbation $\zeta\sim 10^{-5}$ which is ultimately
responsible for the CMB anisotropies. At cosmological horizon re-entry the small-scale fluctuations in the overdensity $\delta\rho$ might 
collapse into a PBH if they are large enough  to overcome the pressure gradients: a PBH would form if the perturbation amplitude $\delta$ 
is larger than a given threshold $ \delta_c$, with a mass  of the order of the mass contained  within the horizon volume at horizon re-entry. 
The mechanism of PBH formation has been investigated in details by several authors performing spherically symmetric numerical simulations 
\cite{Jedamzik:1999am,Shibata:1999zs,Hawke:2002rf,Hawke:2002rf,Musco:2004ak} and it has been shown that the critical collapse mechanism 
\cite{Choptuik:1992jv} arises when $\delta > \delta_c$, with the mass spectrum of PBHs described by a scaling law 
\cite{Niemeyer:1997mt,Musco:2008hv, Musco:2012au}.

The  abundance $\beta$ of PBHs at formation is exponentially  sensitive to the threshold (for simplicity we give the Gaussian expression)
\be
\label{form}
\beta \equiv \left.\frac{\rho_\PBH}{\rho_{\rm tot}}\right|_{\rm form}=P_\G(\delta>\delta_c) = 
\int_{\delta_c}\frac{\d\delta }{\sqrt{2\pi}\sigma}\,e^{-\delta^2/2\sigma^2}\simeq\sqrt{\frac{1}{2\pi}}\frac{\sigma}{\delta_c}e^{-\delta_c^2/2\sigma^2}
\ee
where $\rho_\PBH$ is the energy density collapsed into PBHs while $\rho_{\rm tot}$ is the total energy density. The expression $P_\G(\delta>\delta_c)$ 
indicates the Gaussian probability of a perturbation collapsing to a PBH if its amplitude $\delta$ is larger than a certain threshold $\delta_c$. 
Here $\sigma^2$ is the variance of the overdensity 
\be
\sigma^2 = \int \frac{{\rm d}^3k}{(2\pi)^3}\,W^2(k,R_H)\,P_{\delta}(k),
\ee
where $P_\delta$ is the overdensity power spectrum, $R_H$ being the comoving horizon length  $R_H=1/aH$,  $H$ is the Hubble rate 
and $a$ the scale factor. The quantity $W(k,R_H)$ is an appropriate window function.

 Recently the investigation of the value of the threshold $\delta_c$ has been very intensive and it has been pointed out that the value  of 
$\delta_c$ is not unique, but depends on the shape of the power spectrum of the curvature perturbation \cite{haradath,musco,mg}. In particular 
the exact value, lineraly extrapolated at horizon crossing, is varying between $0.4$ and $2/3$, depending on the particular initial shape of the 
curvature/density profile \cite{musco}, which affects the impact of the pressure gradients during the non-linear evolution of the collapse. 
This is closely related to the shape of the power spectrum which determines, using peak theory, the average shape of the density perturbation 
\cite{BBKS,mg}.

One point of particular importance is the fact that the overdensity\footnote{The notation used here for the density constrast is slightly different 
from what has been used in the literature. Usually papers on PBH formation are using $\delta\rho/\rho_b$ while other papers, coming from the 
cosmological community, use the simpler notation $\delta$ for the same quantity. Because with $\delta$ here we are referring to the average 
threshold integrated over the volume, to keep a clear distinction between the two quantities, we have decided to simplify a bit the notation calling 
the density contrast just as $\delta\rho$, properly defined later in \eqref{rel}.} $\delta\rho$ and the curvature perturbation $\zeta$ are related to 
each other by a non-linear relation. In the comoving slicing, when the Universe is radiation dominated, it reads \cite{harada} 
 \be
 \label{curv_nl}
 \delta\rho(\vec x,t) = -\left(\frac{2\sqrt{2}}{3aH}\right)^2 e^{-5\zeta(\vec x)/2}\nabla^2e^{\zeta(\vec x)/2}.
 \ee
 This implies that, even when the curvature perturbation is a Gaussian random field, the overdensity $\delta\rho$ is intrinsically and unavoidably 
 non-Gaussian \cite{ng2,ng3,ng4}. In the presence of such ineludible non-Gaussianity the abundance of PBHs is significantly reduced compared 
 to the Gaussian (linear) case where one approximates the relation 
 (\ref{curv_nl}) as
\beq\label{curv_l}
\delta\rho(\vec x,t)\simeq - \left (\frac{2}{3aH} \right)^2 \nabla^2\zeta(\vec x) \,,
\ee
and one has to reduce the amplitude of the power spectrum of $\zeta$ by a factor ${\cal O}(2\div 3)$ to have the same non-Gaussian number of 
PBHs starting from the Gaussian expression \cite{ng2,ng3,ng4}. 

The goal of this paper is to assess the impact  of the intrinsic non-Gaussianity of the overdensity  onto the critical threshold $\delta_c$. 
PBHs are identified with the local maxima of the overdensity and,  in order to distinguish whether a cosmological perturbation will collapse forming a 
PBH, it is crucial to evaluate the amplitude of the  peak of the corresponding compaction function. An important input is therefore the shape of the 
overdensity around the peak since non-linearities  would have an effect on the shape of the density, and it is reasonable to expect that the intrinsic 
non-Gaussianity modifies the critical threshold $\delta_c$. As a byproduct, our investigation will allow us to check (and in fact confirm a posteriori) the 
validity of the  assumption made in  \cite{ng3} where the abundance of the PBHs, including the effect coming from the intrinsic non-Gaussianity, has 
been performed adopting the critical threshold $\delta_c$ derived for the linear Gaussian relation between $\delta\rho$ and $\zeta$. 

Our results are based on a perturbative calculation of the average profile around the peak of a perturbation and suggest that the critical threshold is 
rather robust against the intrinsic non-Gaussianity introduced by the non-linear relation between $\delta\rho$ and $\zeta$. This also remains true if 
we endow the curvature perturbation with some primordial non-Gaussianity. The relative changes of the critical threshold are of the order of few percent 
and they do not significantly affect the calculation of the PBH abundance. The reason for this result is  based on the close relation between the shape of 
the density perturbation and the value of the threshold $\delta_c$: although the amplitude of the non-linear components is of the same order of the linear 
one, the effect on the shape due to the non-linear and the non-Gaussian effects are not very significant, and therefore the final shape is quite close to the 
one obtained using the linear approximation given by \eqref{curv_l}. For this reason we suggest, as other works have done (e.g. \cite{ng4,Young:2019osy}), 
that the threshold $\delta_c$ allows the computation of the abundance of PBHs with less uncertainties with respect of using the local critical amplitude of the 
peak which is more sensitive to the local features of the shape.   

The paper is organised as follows. In Section \ref{Initial Conditions} we describe how to specify initial conditions for PBH formation. Section \ref{average profile} 
is devoted to the calculation of the average density profile in the presence of non-Gaussianity. In section \ref{Threshold} we discuss average density profile 
around the threshold for PBH formation, assuming a particular shape of the power spectrum to derive the explicit shape of the density, which is then discussed 
in Section \ref{Averaged profiles}. In Section \ref{Numerical results} we discuss the numerical results obtained with the initial conditions previously derived, and 
finally in Section \ref{Conslusions} we give our conclusions.


\section{Initial conditions for PBH formation}
\label{Initial Conditions}
\noindent
In order to describe the formation of PBHs, we need to consider a region of the expanding Universe with a local non-linear perturbation of the metric which, 
after re-entering the cosmological horizon, will collapse forming a black hole. Assuming spherical symmetry the perturbation of this region is described by the 
two following asymptotic forms of the metric 
\be
\d s^2 = -\d t^2 + a^2(t) \left[ \frac{\d r^2}{1-K(r)r^2} + r^2\d\Omega^2 \right]=-\d t^2+a^2(t)e^{2\zeta(\hr)}\d\vx^2,
\label{pert_metric}
\ee
where the equivalence between the radial and the angular parts gives
\begin{equation} 
\left\{
\begin{aligned}
& r = \hr e^{\zeta(\hr)} \,,\\ 
& \displaystyle{\frac{\d r}{\sqrt{1-K(r)r^2}}} = e^{\zeta(\hr)} \d\hr \,.
\end{aligned}
\right.
\label{K_zeta}
\end{equation}
In Eq. \eqref{pert_metric} $a(t)$ is the scale factor while $K(r)$ and $\zeta(\hr)$ are the conserved comoving curvature perturbations on super-Hubble scale, 
converging to zero at infinity where the universe is taken unperturbed and spatially flat. Combining the two expressions of Eq. \eqref{K_zeta} one gets the 
explicit transformation between $K(r)$ and $\zeta(\hr)$
\be\label{K-zeta}
K(r)\,r^2 = -\hr\zeta'(\hr) \left[ 2+\hr\zeta'(\hr) \right] \,,
\ee
where $\zeta'(\hr)$ is the first derivative of $\zeta(\hr)$ with respect to $\hr$. In general $K(r)$ and $\zeta(\hr)$ are identified with  
the average curvature profile.

The metrics given by Eq. \eqref{pert_metric} are asymptotic solutions of the Einstein equations, while the full solution on superhorizon scales, when the 
curvature profile is conserved being time independent, is obtained using the gradient expansion approximation 
\cite{Shibata:1999zs,Tomita:1975kj,Salopek:1990jq,Polnarev:2006aa}. In this regime the energy density profile can be written as a function of the curvature 
profile \cite{harada,musco} as
\be
\label{rel} 
\delta\rho \equiv \frac{\rho(r,t) - \rho_b(t)}{\rho_b(t)} = \frac{1}{a^2H^2} \frac{(1+\omega)}{5+3\omega} \frac{ \left[ K(r)\,r^3 \right]^\prime}{r^2} 
= - \frac{1}{a^2H^2} \frac{4(1+w)}{5+3w} e^{-5\zeta(\hr)/2}\nabla^2 e^{\zeta(\hr)/2} \,.
\ee
Here $H(t) \equiv \dot{a}(t)/a(t)$ is  the Hubble parameter while $\omega$ is the coefficient of the equation of state relating the total (isotropic) pressure $p$ 
to the total energy density $\rho$ as
\be
p = \omega \rho \,,
\ee
where the standard scenario for PBHs assumes a radiation dominated Universe with $\omega=1/3$. The difference between the two Lagrangian coordinates 
$r$ and $\hr$ is related to the particular parameterization of the comovingcoordinate, fixed by the particular form chosen to specify the curvature perturbation 
into the metric, i.e. $K(r)$ or $\zeta(\hr)$.Here $K'(r)$ denotes differentiation with respect to $r$ while $\zeta'(\hr)$ and $\nabla^2\zeta(\hr)$ denote differentiation 
with respect to $\hr$.

The criterion to distinguish whether a cosmological perturbation is able to form a PBH depends on the amplitude measured at the peak of the compaction 
function\footnote{This was originally introduced in \cite{Shibata:1999zs} without the factor 2.} defined as
\be
\label{a}
\mathcal{C} \equiv 2\frac{\delta M(r,t)}{R(r,t)} \,,
\ee
where $R(r,t)$ is the areal radius 
and $\delta M(r,t)$ is the difference between the Misner-Sharp mass within a sphere of  radius $R$ and background mass $M_b(r,t)=4\pi \rho_b(r,t)R^3(r,t)/3$ 
with the same areal radius but calculated with respect to a spatially  flat FRW metric. In the superhorizon regime, applying the gradient expansion approximation, 
the compaction function is time independent, and is simply related to the curvature profile by
\be
\mathcal{C} = \frac{3(1+w)}{5+3w} K(r)r^2 \,, 
\ee
which,  using Eq. \eqref{K-zeta},  can be written also in terms of $\zeta(\hr)$. As shown in \cite{musco}, the length-scale of the perturbation must be identified as 
the location where the compaction function is reaching its maximum
\be
\mathcal{C}'(r_m) = 0 \quad \textrm{or} \quad \mathcal{C}'(\hr_m) = 0
\ee
which gives
\be \label{eq_rm}
K(r_m)+\frac{r_m}{2}K'(r_m)=0 \quad {\rm or} \quad \zeta'(\hr_m)+\hr_m\zeta''(\hr_m)=0 \,.
\ee
Given the curvature profile, the value of $r_m$ or $\hr_m$ can be then used to define the small parameter $\epsilon$ of the gradient expansion approximation as
\be
\epsilon \equiv \frac{R_H(t)}{R_b(r_m,t)} = \frac{1}{aHr_m} = \frac{1}{aH\hr_m e^{\zeta(\hr_m)}} \,,
\ee
where $R_H$ is the cosmological horizon and $R_b(r,t)=a(t)r$ is the areal radius of the background (note that in terms of $\hr_m$ the curvature profile $\zeta(\hr_m)$ 
is necessary to compute the background value of the areal radius, because of the difference between $r$ and $\hr$). The explicit form of the density profile seen in 
Eq. \eqref{rel}, valid for small $\epsilon$, is given by
\be \label{deltarho_zeta}
\delta\rho = \left( \frac{1}{aH} \right)^2 \frac{3(1+\omega)}{5+3\omega} \left(K(r)+\frac{r}{3}K'(r)\right) = 
- \left( \frac{1}{aH} \right)^2 \frac{2(1+\omega)}{5+3\omega} 
\left[ \zeta''(\hr) + \zeta'(\hr)\left(\frac{2}{\hr} + \frac{1}{2}\zeta'(\hr)\right) \right] e^{-2\zeta(\hr)} \,,
\ee
where in the first equality the term $[K(r)r^3]'$ of Eq. (\ref{rel}) has been written explicitly, while in the second equality 
$\nabla^2 e^{\zeta(\hr)/2}$ has been written in spherical symmetry. Note that to write explicitly these expressions in terms of 
the small parameter $\epsilon$ one needs to insert $r_m$ into the denominator of the term $(1/aH)$ and multiply the radial profile
by $r_m^2$.  

Introducing only a perturbation of the energy density field as initial condition corresponds to a combination of growing and decaying 
mode which would affect the evolution of the cosmological perturbation and the corresponding value of the threshold. As noticed also 
in \cite{Musco:2004ak,musco}, to have at initial time a perturbation behaving like a pure growing mode it is necessary to introduce also 
a consistent perturbation of  the velocity field $U$ and the areal radius $R$, that in gradient expansion have the following form
\begin{eqnarray}
& & U = H(t)R ( 1 +  \delta U ) \label{U_pert},  \\
& & R = a(t)r ( 1 + \delta R ) \label{R_pert}, 
\end{eqnarray}
where for a pure growing mode one has
\begin{eqnarray}
& & \delta U = - \frac{1}{(1+\omega)} \frac{1}{r^3} \int \, r^2  \,\d r  \,\delta\rho \label{delta_U}, \\
& & \delta R = - \frac{\omega}{(1+3\omega)(1+\omega)} \delta\rho + \frac{1}{1+3w} \delta U   \label{delta_R}.
\end{eqnarray}
We are now able to define consistently the perturbation amplitude as the mass excess of the energy density within the scale $r_m$
measured at horizon crossing time $t_H$, defined when $\epsilon=1$ ($aHr_m=1$). Although in this regime the gradient 
expansion approximation is not very accurate and the perturbation amplitude does not represent the exact value of the perturbation at the 
``real horizon crossing", this provides a well defined criterion that allows one to compare consistently the amplitude of different 
perturbations, understanding how the threshold is varying because of the different initial curvature profiles (see \cite{musco} 
for more details). The amplitude of the perturbation is given by the excess of mass averaged over a spherical volume of radius $R_m$, 
defined as
\be
\label{delta}
\delta(r_m,t_H) = \frac{4\pi}{V_{R_m}} \int_0^{R_m} \, \d R\, R^2  \delta\rho  = 
\frac{3}{r_m^3} \int_0^{r_m} \,  \d r\,r^2 \,\delta\rho \,, \quad \textrm{where} \quad
\quad V_{R_m} = \frac{4\pi}{3}R_m^3.
\ee
The second equality is obtained by neglecting the higher order terms in $\epsilon$, which allows $R(r,t)$ to be approximated as \mbox{$R(r,t) 
\simeq a(t)r$}, reducing the first integral over the physical sphere of areal radius $R_m$ to an integral over the comoving 
volume of radius $r_m$. inserting the explicit expression of $\delta\rho$ in terms of the curvature profile into \eqref{delta} 
one obtains
\be\label{delta_m}
\delta_m \equiv \delta(r_m,t_H) = \frac{3(1+w)}{5+3w} K(r_m)r_m^2 = \mathcal{C}(r_m)
\ee
and a simple calculation seen in \cite{musco} gives the fundamental relation
\be
\delta_m = 3 \delta\rho (r_m,t_H) \,. 
 \label{delta_m}
\ee
Inserting now Eq. \eqref{delta_m} into Eq. \eqref{delta} one can easily show that
\be\label{r_m}
r_m^3 = \frac{\displaystyle{\int_0^{r_m} \, \d r\,r^2  \, \delta\rho(r,t_H) } }{\delta\rho(r_m,t_H)} \,,
\ee
which gives an alternative way to compute the length scale $r_m$ of the perturbation directly from the energy density 
profile instead of using the curvature profile.

As shown in \cite{musco} the threshold of $\delta_m$ for PBH formation, called $\delta_c$, depends crucially on the shape of the 
perturbation, which we parameterize in the following through the average density contrast $\overline{\delta\rho}(r)$ measured 
at horizon crossing $t_H$. This quantity inevitably receives non-Gaussian corrections, even though the comoving curvature 
perturbation is Gaussian. This because the relation (\ref{rel}) between the density contrast $\delta\rho$ and the comoving curvature 
perturbation $\zeta$ is non-linear. In the next section we will calculate the average density contrast $\overline{\delta\rho}(r)$ 
away from a threshold in the presence of non-Gaussianity.


\section{The average density profile} 
\label{average profile}
\noindent
To the best of our knowledge, the  average profile around a peak has not been calculated for the non-Gaussian case in peak theory. 
We will therefore resort to threshold statistics, reviewing first the calculation for the Gaussian case in Section \ref{Gaussian}, generalizing 
it for the non-Gaussian case then in Section \ref{non-Gaussian}. Since regions  with  peak amplitude 
$\delta\rho_0\gg  \sigma$  
 correspond  to local maxima  to high statistically degree \cite{hoffman,ng3},  our approach should be enough when dealing with PBHs.

\subsection{The Gaussian case} 
\label{Gaussian}
\noindent
Let us first recall how to compute for a Gaussian statistics
 the average profile $\overline{\delta\rho}(x_1)$ of the density contrast $\delta\rho(\vx_1)$
at a given point $\vx_1$ from a threshold point located at 
  $\vx_2$  \cite{pw}. 
We define  the distance $|\vx_2-\vx_1|=r$.
 Assuming spherical symmetry we can  write $\delta\rho(\vx_1)=\delta\rho(r)$ and   $\delta\rho(\vx_2)=\delta\rho_0>\nu\sigma$, 
 where $\sigma^2=\langle\delta\rho^2(\vx)\rangle$ is the variance of the density contrast.

At a  distance $r$ from the threshold at the origin, the average $\delta\rho$ is given by 
\be
\label{aa}
\overline{\delta\rho}(r)=\langle \delta\rho(r)|\delta\rho_0>\nu\sigma\rangle=\int_{-\infty}^{\infty}{\rm d}\delta\rho(r) \,\delta\rho(r) P(\delta(r)|\delta\rho_0>\nu\sigma), 
\ee
where
\begin{eqnarray}
P(\delta\rho(r)|\delta\rho_0>\nu\sigma)&=&\frac{P(\delta\rho(r),\delta\rho_0>\nu\sigma)}{P(\delta\rho_0>\nu\sigma)}.
\end{eqnarray}
Since $\delta\rho(r)$ and $\delta\rho_0$ are Gaussian variables, one  can derive 
$P(\delta\rho(r),\delta\rho_0)$ using  the covariance matrix
\begin{eqnarray}
P(\delta\rho(r),\delta\rho_0)&=&\frac{1}{2\pi\sqrt{\det C}}\exp\left(-\vec{\delta\rho}^T C^{-1}\vec\delta\rho/2\right)\,\nonumber\\
\vec{\delta\rho}^T&=&(\delta\rho_0,\delta\rho(r)),\nonumber\\
C&=&\left(\begin{array}{cc}
\sigma^2&\xi^{(2)}(r)\\
\xi^{(2)}(r) & \sigma^2\end{array}\right),
\end{eqnarray}
where
\be
\xi^{(2)}(r)=\langle\delta\rho(\vx_1)\delta\rho(\vec 0)\rangle
\ee
denotes  the two-point correlator. We then deduce that
\begin{eqnarray}
P(\delta\rho(r),\delta\rho_0>\nu\sigma)&=&\frac{e^{-\delta\rho^2(r)/2\sigma^2}}{2\sqrt{2\pi}\sigma}\left(1+{\rm Erf}\left[\frac{\left(\xi^{(2)}(r)\delta\rho(r)-\nu\sigma^3\right)}{\sigma\sqrt{2\det C}}\right]\right),\nonumber\\
P(\delta\rho_0>\nu\sigma)&=&\frac{1}{2}{\rm Erfc}\left(\nu/\sqrt{2}\right),
\end{eqnarray}
being  ${\rm Erfc}(x)$  the complementary error function. Using Eq. (\ref{aa}), we then obtain
\be
\label{fg}
\overline{\delta\rho}(r)=\frac{\xi^{(2)}(r)}{\sigma}\sqrt{\frac{2}{\pi}}\frac{e^{-\nu^2/2}}{{\rm Erfc}\left(\nu/\sqrt{2}\right)}.
\ee
Finally, exploiting  the asymptotic behaviour
\be
{\rm Erfc}\left(x\gg 1\right)\approx \frac{e^{-x^2}}{x\sqrt{\pi}},
\ee
we get that the  average $\overline{\delta\rho}$ at a distance $r$ from  the threshold with  $\nu\gg 1$ is
\be
\overline{\delta\rho}(r)\simeq \nu \,\frac{\xi^{(2)}(r)}{\sigma}.
\ee
As expected, for large values of $\nu$, it coincides with the average profile around peaks obtained  using peak theory \cite{bbks}.


\subsection{The non-Gaussian case} 
\label{non-Gaussian}
\noindent
In this section we generalise the calculation of the average density profile to the case in which the density contrast is a non-Gaussian field. 
We start by defining more conveniently the probability
\be
P(\delta\rho(r),\delta\rho_0>\nu\sigma)=\Big<\delta_D(\delta\rho-\delta\rho(r))\theta(\delta\rho_0-\nu\sigma)\Big>,
\ee
where $\theta(x)$ is the standard step function, and the conditional probability is therefore
\begin{eqnarray}
P(\delta\rho(r)|\delta\rho_0>\nu\sigma)&=&\frac{\Big<\delta_D(\delta\rho-\delta\rho(r))\theta(\delta\rho_0-\nu\sigma)\Big>}{\Big<\theta(\delta\rho_0-\nu\sigma)\Big>}.
\end{eqnarray}
To proceed, we closely   follow
the path-integral technique developed in   \cite{blm,pbhng}.
Our starting point is the   density contrast   $\e(\vx)$  endowed with a  probability distribution $P[\e(\vx)]$.  The corresponding  partition function 
$Z[J]$ in the presence of an external source $J(\vx)$ reads
\begin{eqnarray}
\label{part}
 Z[J]=\int [{D}\e(\vx)]P[\e(\vx)]e^{i\int \d^3 x J(\vx)\e(\vx)},\quad \int [D\e(\vx)]P[\e(\vx)]=1. 
 \end{eqnarray} 
The  connected $n$-point correlation functions are generated by the  functional Taylor expansion of $W[J]=\ln Z[J]$  in powers of the source $J(\vx)$  
\be
\xi^{(n)}=\xi^{(n)}(\vx_1,\cdots,\vx_n)=\langle\delta(\vx_1),\cdots,\delta(\vx_n)\rangle_c.
\ee
At this stage, it is also convenient to  normalise the  correlators
as  
\begin{eqnarray}
w^{(n)}(\vx_1,\cdots,\vx_n)=\sigma^{-n}\, \xi^{(n)}(\vx_1,\cdots,\vx_n).
\end{eqnarray}
For instance, 
\begin{eqnarray}
w^{(2)}(0)=1
\end{eqnarray}
denotes the two-point correlator evaluated at the same point.

With our formalism the  average density contrast is easily found as 
\begin{eqnarray}
\label{sim}
\overline{\delta\rho}(r)&=&\langle \delta\rho(r)|\delta\rho_0>\nu\sigma\rangle =
\int_{-\infty}^{\infty}{\rm d}\delta\rho(r) \,\delta\rho(r)
\frac{P(\delta\rho(r),\delta\rho_0>\nu\sigma)}{P(\delta\rho_0>\nu\sigma)},\nonumber\\
&=& \frac{1}{P(\delta\rho_0>\nu\sigma)}\int_{-\infty}^{\infty}{\rm
d}\delta\rho(r) \,\delta\rho(r)
\Big<\delta_D(\delta\rho(\vx)-\delta\rho(r))\theta(\delta\rho_0-\nu\sigma)\Big>\nonumber\\
&=& \frac{1}{P(\delta\rho_0>\nu\sigma)}\int_{-\infty}^{\infty}{\rm
d}\delta\rho(r) \,\delta\rho(r) \int [D\delta\rho(\vx)]P[\delta\rho(\vx)]
\delta_D(\delta\rho(\vx)-\delta\rho(r))\theta(\delta\rho_0-\nu\sigma)\nonumber \\
&=&\frac{1}{P(\delta\rho_0>\nu\sigma)} \int [D\delta\rho(\vx)]P[\delta\rho(\vx)]
\, \delta\rho(\vx) \,  \theta(\delta\rho_0-\nu\sigma)=\frac{\Big<\delta\rho(\vx)\,
\theta(\delta\rho_0-\nu \sigma)\Big>}{\Big<\theta(\delta\rho_0-\nu\sigma)\Big>}.
\end{eqnarray}
To evaluate it, we use the following representation of the $\theta$-function
\begin{eqnarray}
\theta(x)=\int_{-x}^\infty {\rm d}a \int_{-\infty}^\infty \frac{{\rm d} \phi}{2\pi}
e^{i\phi a}, 
\end{eqnarray}
and the identity 
\begin{eqnarray}
x=\int_{-\infty}^\infty {\rm d} a\, a\, \delta_D(a-x) =
\int_{-\infty}^\infty {\rm d} a \, a  \int_{-\infty}^\infty \frac{{\rm d} \phi}{2\pi}
e^{i\phi (a-x)}. 
\end{eqnarray}
This implies
\begin{eqnarray}
\Big<\delta\rho(\vx_1)\, \theta(\delta\rho(\vx_2)-\nu \sigma)\Big>=(2\pi)^{-2}\sigma
\int_{-\infty}^\infty {\rm d} a_1\, a_1\int_{\nu}^\infty
{\rm d} a_2\int_{-\infty}^\infty{\rm d} \phi_1 \int_{-\infty}^\infty
{\rm d} \phi_2 e^{-i\sigma(\phi_1a_1+\phi_2 a_2)}Z[J], \label{dd}
\end{eqnarray}
with
\begin{eqnarray}
J(\vx)=\phi_1 W(|\vx-\vx_1|,R)+\phi_2 W(|\vx-\vx_2|,R).
\end{eqnarray} 
Using the standard expansion for  $\ln Z[J]$ 
\begin{align*}
\ln Z[J]&= \sum_{n=2}^\infty\frac{i^n}{n!}\int \d^3 \vy_1\cdots \int \d^3 \vy_n \,
\sum_{i_1=1}^N\cdots \sum_{i_n=1}^N J_{i_1}(\vy_1,\vx_1)\cdots J_{i_n}(\vy_n,\vx_n)
 \xi^{(n)}(\vy_1,\cdots,\vy_n) \\
 & = \sum_{n=2}^\infty\frac{i^n}{n!} 
 \sum_{m=0}^n{ {n}\choose{m}} \phi_1^m \phi_2^{n-m}
\xi^{(n)}_{R;[m,n-m]},
\end{align*}
where 
\begin{equation}
\xi^{(n)}_{R;[m,n-m]} =
\xi^{(n)}_{R}\big(
\underbrace{\vx_1,\cdots,\vx_1}_{m\text{-times}},\underbrace{\vx_2,\cdots,\vx_2}_{(n-m)\text{-times}}\big) ,
\end{equation}
we find
\begin{align}
\Big<\delta\rho(\vx_1)\, \theta(\delta\rho(\vx_2)-\nu \sigma)\Big>&=(2\pi)^{-2}\sigma
\int_{-\infty}^\infty {\rm d} a_1\, a_1\int_{\nu}^\infty
{\rm d} a_2\int_{-\infty}^\infty{\rm d} \phi_1 \int_{-\infty}^\infty
{\rm d} \phi_2 \nonumber \\
&\exp\left\{  \sum_{n=2}^\infty\frac{i^n}{n!} 
 {\sum_{m=0}^n}'{ {n}\choose{m}} i^n
\xi^{(n)}_{R;[m,n-m]}\frac{\partial^m}{\partial a_1^m}
\frac{\partial^{n-m}}{\partial a_2^{n-m}}
\right\}\nonumber \\
&\exp{\left(-\frac{1}{2}\sigma^2(\phi_1^2+\phi_2^2)-
i\sigma(\phi_1a_1+\phi_2a_2)\right)}.
\end{align}
Here the  prime on the sum reminds us  that the sum has to be performed by omitting  the terms containing $\phi_1^2$ and $\phi_2^2$, and  
performing the integration over the variables $\phi_1$ and $\phi_2$ 
\begin{align}
\Big<\delta\rho(\vx_1)\, \theta(\delta\rho(\vx_2)-\nu \sigma)\Big>& = (2\pi)^{-1}\sigma
\int_{-\infty}^\infty {\rm d} a_1\, a_1\int_{\nu}^\infty
{\rm d} a_2  \nonumber \\
&\exp\left\{  \sum_{n=2}^\infty\frac{(-1)^n}{n!} 
 {\sum_{m=0}^n}'{ {n}\choose{m}} 
w^{(n)}_{R;(m,n-m)}\frac{\partial^m}{\partial a_1^m}
\frac{\partial^{n-m}}{\partial a_2^{n-m}}
\right\}\exp\left(-\frac{1}{2}(a_1^2+a_2^2)\right) , \label{ss0}
\end{align}
 we find
 \begin{eqnarray}
\Big<\delta\rho(\vx_1)\, \theta(\delta\rho(\vx_2)-\nu \sigma)\Big>&=
\frac{\sigma}{\sqrt{2\pi}}e^{-\nu^2/2}\left(w^{(2)}(\vx_1,\vx_2)+
\frac{\nu}{2} w^{(3)}(\vx_1,\vx_2,\vx_2)+\frac{\nu^2-1}{6}
w^{(4)}(\vx_1,\vx_2,\vx_2,\vx_2)+\cdots\right). \label{ss1}
\end{eqnarray}
It is relevant to point out that only connected correlators appear up to fourth-order, whereas 
non-connected correlators start to appear at the next fifth order. Then, the  connected piece  of (\ref{ss1}) turns out to be  
\begin{eqnarray}
\Big<\delta\rho(\vx_1)\, \theta(\delta\rho(\vx_2)-\nu \sigma)\Big>_{c}
=\frac{\sigma}{\sqrt{2\pi}}e^{-\nu^2/2}\sum_{m=0}^\infty \frac{1}{2^{m/2}(m+1)!}
w^{(m+2)}(\vx_1,\vx_2,\cdots,\vx_2)H_{m}\left(\frac{\nu}{\sqrt{2}}\right),
\end{eqnarray}
where $H_m(x)$ are the Hermite polynomials. 

The one-point non-Gaussian threshold probability   for $\nu\gg 1$  is \cite{blm,pbhng}
\be
\Big<\theta(\delta\rho_0-\nu\sigma)\Big>\approx \frac{e^{-\nu^2/2}}{\sqrt{2\pi}\nu}{\rm exp}\left(\sum_{n=3}^\infty \nu^n w^{(n)}(0)/n!\right),
\ee
where  $w^{(n)}(0)$ are the normalised $n$-point correlators calculated at the same point.  The final expression of the average profile at distance $r$ from the 
origin and for large thresholds  and up to the four-point correlator  (recall that $|\vx_2-\vx_1|=r$) is given by
\begin{eqnarray}
\label{final}
\overline{\delta\rho}(r)=\nu\left[\frac{\xi^{(2)}(r)}{\sigma}+
\frac{\nu}{2\sigma^2} \xi^{(3)}(\vx_1,\vx_2,\vx_2)+\frac{\nu^2}{6\sigma^3}
\xi^{(4)}(\vx_1,\vx_2,\vx_2,\vx_2)+\cdots\right]{\rm exp}\left(-\sum_{n=3}^\infty (\nu/\sigma)^n \xi^{(n)}(0)/n!\right).
\end{eqnarray}
Of course  it reduces to the expression (\ref{fg}) once the Gaussian limit is taken. The expression (\ref{final}) is the profile we are going to use in the following. 
However, we will restrict ourselves  to a perturbative approach and only include  the three-point correlator. Including higher-order terms is unfortunately technically 
quite demanding. However we will show in Section \ref{Numerical results}  that the modifications of the threshold due to the three-point correlator is quite small 
because the final non-Gaussian shape is not very different with respect the linear Gaussian one.


\section{The average density profile around the threshold for PBH formation} 
\label{Threshold}
\noindent
Having calculated the generic expression for the average profile around the threshold, we are now ready to study the problem of PBH formation. As we  
have already stressed, equation (\ref{rel}) is a non-linear relation between the density contrast $\delta\rho$ and the comoving curvature perturbation $\zeta$. 
This makes the variable $\delta\rho$ non-Gaussian even if $\zeta$ is Gaussian. 

First, we will assume that the comoving curvature $\zeta$ is Gaussian so that  $\zeta$  does not have an intrinsically second-order component $\zeta_2$, 
but only the linear one, which will call $\zeta_1$  (we will promptly extend our computation
to the case in which $\zeta$ has some primordial non-Gaussianity). 
Furthermore, we will restrict ourselves to the case in which we keep only the three-point correlator (see more comments on this  later on). Let us also 
notice that the  description of the PBH collapse involves a non-linear relation between the areal radius $R$ and the comoving coordinate $r$, i.e. 
$R=r \,{\rm exp}\,\zeta$, which introduces further non-linearities. 

Expanded at second-order for a linear Gaussian comoving curvature pertubation $\zeta_1$, the density contrast is made of  a first- and a second-order 
piece (we assume from now on a radiation phase)
\begin{eqnarray}
\delta\rho&=&\delta\rho_1+\delta\rho_2,\nonumber\\
\delta\rho_1&=&\frac{4}{9}\frac{1}{a^2H^2}\nabla^2\zeta_1,\nonumber\\
\delta\rho_2&=&-\frac{8}{9}\frac{1}{a^2H^2}\left(\frac{1}{4}(\nabla\zeta_1)^2-\zeta_1\nabla^2\zeta_1\right). 
\label{delta_secondorder}
\end{eqnarray}
In Fourier space these relations become (we use here the conventions of  \cite{b})
\begin{eqnarray}
\delta\rho_1(\vk)&=&\alpha(k)\zeta_1(\vk),\,\,\,\,\alpha(k)=-\frac{4}{9}\frac{k^2}{a^2H^2}\nonumber\\
\delta\rho_2(\vk)&=&\int\d^3 k_1\int\d^3 k_2\,\delta(\vk-\vk_1-\vk_2)\,F(\vk_1,\vk_2)\,\zeta_1(\vk_1)\zeta_1(\vk_2) 
\label{d12}
\end{eqnarray}
where
\[ F(\vk_1,\vk_2)=\frac{8}{9}\frac{1}{a^2H^2}\left(\frac{1}{4}\vk_1\cdot\vk_2-\frac{1}{2}(k_1^2+k_2^2)\right). \]
The corresponding bispectrum turns out to be
\begin{eqnarray}
\Big< \delta\rho_1(\vk_1) \delta\rho_1(\vk_2) \delta\rho_2(\vk_1)\Big>&=&\delta(\vk_1+\vk_2+\vk_3)\,B_\zeta(\vk_1,\vk_2,\vk_3),\nonumber\\
B_\zeta(\vk_1,\vk_2,\vk_3)&=&2\alpha(k_1)\,\alpha(k_2)\,F(\vk_1,\vk_2)P_{\zeta_1}(k_1)P_{\zeta_1}(k_2)+{\rm cyclic}\nonumber\\
&=&2\frac{F(\vk_1,\vk_2)}{\alpha(k_1)\,\alpha(k_2)}P_{\delta\rho_1}(k_1)P_{\delta\rho_1}(k_2)+{\rm cyclic},
\end{eqnarray}
where $P_{\zeta_1}(k)$ and $P_{\delta\rho_1}(k_2)$ are the power spectrum of the comoving curvature perturbation and of the linear density contrast, 
respectively. The connected two-point and three-point correlators in coordinate space are given by
\begin{eqnarray}
\xi^{(2)}(\vx_1,\vx_2)&=&\int\d^3 k\,e^{i\vk\cdot\vx}\,P_{\delta\rho_1}(k)=4\pi\int\d k \,k^2\,\frac{\sin kr}{kr}\,P_{\delta\rho_1}(k)=4\pi\int\d k \,k^2\,\frac{\sin kr}{kr}\,\alpha^2(k)P_{\zeta}(k)
\end{eqnarray}
and
\begin{eqnarray}
\xi^{(3)}(\vx_1,\vx_2,\vx_3)&=&\Big< \delta\rho_1(\vx_1) \delta\rho_1(\vx_2) \delta\rho_3(\vx_2)\Big>_c=\int\d^3 k_1\int\d^3 k_2\int\d^3 k_3\,e^{i\vk_1\cdot \vx_1+i\vk_2\cdot \vx_2+i\vk_3\cdot \vx_3}\,\Big< \delta\rho_1(\vk_1) \delta\rho_1(\vk_2) \delta\rho_3(\vk_1)\Big>\nonumber\\
&=&\int\d^3 k_1\int\d^3 k_2\int\d^3 k_3\,e^{i\vk_1\cdot \vx_1+i\vk_2\cdot \vx_2+i\vk_3\cdot \vx_3}\,\delta^{(3)}(\vk_1+\vk_2+\vk_3)\,B_\zeta(\vk_1,\vk_2,\vk_3)\nonumber\\
&=&\int\d^3 k_1\int\d^3 k_2\,e^{i\vk_1\cdot(\vx_1-\vx_3)+i\vk_2\cdot (\vx_2-\vx_3)}\,B_\zeta(\vk_1,\vk_2,-\vk_1-\vk_2)
\end{eqnarray}
so that 
\begin{eqnarray}
\xi^{(3)}(\vx_1,\vx_2,\vx_2)&=&\int\d^3 k_1\int\d^3 k_2\,e^{i\vk_1\cdot(\vx_1-\vx_2)}\,B_\zeta(\vk_1,\vk_2,-\vk_1-\vk_2).
\end{eqnarray}

\subsection{The case of a peaked power spectrum}
\label{Peaked}
\noindent
In order to present analytical formulae we adopt the simplest  power spectrum of the comoving curvature perturbation, corresponding to the Dirac-delta case 
\be
\label{dirac}
P_\zeta=\frac{A}{k^2}\delta_D(k-k_*)
\ee
for which  we have
\begin{eqnarray}
P_{\delta_1}=\alpha^2(k_*)P_\zeta(k)&=&\frac{16}{81}\frac{k_*^2}{a^4 H^4} A\,\delta_D(k-k_*),\nonumber\\
 \label{two}
\xi^{(2)}(r)&=&(4\pi)\alpha^2(k_*) \, A\,\frac{\sin k_*r}{k_*r}, \nonumber\\
\xi^{(2)}(0)&=&\sigma^2=(4\pi)\alpha^2(k_*) A,\\
\xi^{(3)}(\vx_1,\vx_2,\vx_2)&=&\frac{4}{\alpha(k_*)}\sigma^4\,\left[2\frac{\sin k_*r}{k_*r}+\frac{1}{8 k_*^4 r^4}\left(1+5 k_*^2 r^2-(1+3 k_*^2 r^2)\cos 2k_*r -2 k_* r \sin 2k_*r\right)\right],\nonumber\\
\xi^{(3)}(0)&=&\frac{12}{\alpha(k_*)}\sigma^4.
\label{three}
\end{eqnarray}
The power spectrum (\ref{dirac}) should be regarded as the limit of zero width of a  more physical power spectrum \cite{Byrnes}.

\subsection{Including non-Gaussianity of the power spectrum}
\label{non-Gaussian spectrum}
\noindent
We can generalise these findings to the case in which the  comoving curvature perturbation is non-Gaussian \cite{ngreview} and we standardly 
parametrise the non-linearities as
\begin{eqnarray}
\label{zeta_2}
\zeta_2=\zeta_1+\frac{3}{5}f_\NL \zeta_1^2.
\end{eqnarray}
This expression  is intended to parametrise the non-linearities  which arise at small scales around the scale $k_*$ \cite{b1,b2,b3,b4}.  We are going 
to consider  both positive and negative values of $f_\NL$,  keeping in mind that positive values evade observational constraints which place an upper 
bound on the allowed amplitude of the primordial power spectrum, and allow a cosmologically relevant population of PBHs on the relevant 
scales\footnote{Positive values of the non Gaussianitiy reduce the variance of the curvature perturbation and the  bound from the second-order gravitational 
waves is relaxed,  while the contrary is happening for negative values \cite{revPBH}}. The corresponding contribution  to the bispectrum is 
\begin{eqnarray}
B^\NL_\zeta(\vk_1,\vk_2,\vk_3)=\frac{3}{5}\cdot 2 f_\NL \Big(P_\zeta(k_1)P_\zeta(k_2)+{\rm cyclic}\Big).
\end{eqnarray}
Since $B^\NL_\delta(\vk_1,\vk_2,\vk_3)=\alpha(k_1)\alpha(k_2)\alpha(k_3)B^\NL_\zeta(\vk_1,\vk_2,\vk_3)$, we have
\begin{eqnarray}
\xi^{(3)}_\NL(\vx_1,\vx_2,\vx_3)&=&\int\d^3 k_1\int\d^3 k_2\int\d^3 k_3\,e^{i\vk_1\cdot \vx_1+i\vk_2\cdot \vx_2+i\vk_3\cdot \vx_3}\,\delta(\vk_1+\vk_2+\vk_3)\,B_\delta(\vk_1,\vk_2,\vk_3)\nonumber \\
&=&2f_\NL\int\d^3 k_1\int\d^3 k_2 
\,e^{i\vk_1\cdot (\vx_1-\vx_3)+i\vk_2\cdot (\vx_2-\vx_3)}\,\delta(\vk_1+\vk_2+\vk_3) \nonumber \\
&\times&\alpha(k_1)\alpha(k_2) \alpha(-|\vk_1+
\vk_2|)P_\zeta(k_1)P_\zeta(k_2)+{\rm cyclic} \,,
\end{eqnarray}
and for a peaked power spectrum of the form (\ref{dirac}) we finally get 
\begin{eqnarray}\label{threeNL}
\xi^{(3)}_\NL(\vx_1,\vx_2,\vx_2)&=&\frac{3}{5}\cdot\frac{4f_\NL}{\alpha(k_*)}
\sigma^4 \left\{\frac{\sin k_* r}{k_* r}-\frac{1}{k_*^4 r^4}\Big[1-(1-2k_*^2 r^2)\cos 2 k_* r-2 k_* r \sin 2k_* r\Big]\right\}, \nonumber\\
\xi^{(3)}_\NL(0)&=&\frac{3}{5}\cdot\frac{12f_\NL}{\alpha(k_*)}\sigma^4.
\end{eqnarray}

\section{The average profile including the  three-point correlation function}
\label{Averaged profiles}
\noindent
In the previous section we have derived the general form of the three-point correlation function related to the non-linear component of the curvature
profile given by \eqref{rel} and to the possible non-Gaussian component of the curvature power spectrum ($f_\NL\neq0$), considering the particular 
case of a peaked power spectrum, which allows to get an analytic solution. In the first part of this section we are going to analyze the energy density 
profile obtained when the three-point correlation function term is taken into account. Although this is just a particular example, it is nevertheless interesting, 
as a matter of principle, to investigate this case, computing the modification obtained on the threshold $\delta_c$ for PBH formation, to get a hint about the
general effect of the non-linearities. 

In the second part of this section we are going to analyze the energy density profile obtained from the averaged profile of the curvature 
perturbation $\overline\zeta$ of a peaked power spectrum if peak theory is applied to $\zeta$ instead of $\delta\rho$ as was done in \cite{haradath}. 
The aim is to make a comparison of the threshold with the profile obtained with threshold statistics, showing that the energy density profile as follows 
from \eqref{rel}, using the averaged curvature profile $\overline\zeta$,  is very different in general from the mean profile. In other words,
the knowledge of $\overline\zeta$ does not give a direct way to compute the corresponding threshold. A  non-Gaussian method to
generalize peak theory, as the one we are using here, is necessary to compute precisely the threshold of PBH formation. 

\subsection{The averaged density profile from threshold statistics}
\label{Averaged density}
\noindent
Considering \eqref{final} up to the three-point correlation function for the power spectrum given by \eqref{dirac} in spherical symmetry 
and inserting Eqs. \eqref{two}, \eqref{three}, \eqref{threeNL}, one obtains the explicit form of the averaged 
density profile given by
\be 
\overline{\delta\rho}(\hat{x}) = \sigma\nu \left[ \frac{\sin\hat{x}}{\hat{x}} + 12\sqrt{\pi A}\nu \left( {\cal F}_1({\hat x}) + 
\frac{3}{5}f_\NL{\cal F}_2({\hat x}) \right) \right] \exp{\left[-4\left( 1+\frac{3}{5}f_\NL \right) \sqrt{\pi A}\nu^3\right]},
\label{rho_NL}
\ee 
where $\hat{x} \equiv k_* \hat{r}$. Note that in \eqref{two} $\xi^{(2)}(r) \propto \sigma^2$ while in \eqref{three} $\xi^{(3)}(r) \propto \sigma^4$, 
therefore the expansion parameter $\nu$ of \eqref{final} corresponds in the explicit profile given by \eqref{rho_NL} to an expansion around the 
peak amplitude of the energy density. The functions ${\cal F}_1({\hat x})$ and ${\cal F}_2({\hat x})$ are modifications of the profile coming from 
the three-point correlation function related respectively to the 
non-linear term of \eqref{delta_secondorder}, and to the non-Gaussianity introduced in \eqref{zeta_2}
These two functions read as
 \begin{eqnarray} 
{\cal F}_1({\hat x}) & = & \frac{2}{3} \left[ \frac{\sin\hat{x}}{\hat{x}} + \frac{ 1 + 5\hat{x}^2  - (1+3\hat{x}^2)\cos{2\hat{x}} - 
2\hat{x}\sin{2\hat{x}}}{16\hat{x}^4} \right] \,, \\ \nonumber \\
{\cal F}_2({\hat x}) & = &  \frac{1}{3} \left[ \frac{\sin\hat{x}}{\hat{x}} - \frac{ 1 - 2\hat{x}\sin{2\hat{x}} - (1-2\hat{x}^2)\cos{2\hat{x}}}{\hat{x}^4} \right],
\end{eqnarray}
where they have been normalized such that ${\cal F}_1(0) = {\cal F}_2(0) = 1$. Note that in the linear limit of a Gaussian density contrast, 
$\xi^{(3)}(\hat{x})=0$ and the profile is simply reduced to the sync function as it has been obtained in Ref. \cite{musco}. Using now \eqref{d12} 
combined with \eqref{two}, one gets
\be 
4\sqrt{\pi A}\,\nu = \frac{9}{2}\frac{\delta\rho_{0_G}}{x^2_{m_G}} , \quad\quad \textrm{where} \quad\quad x_{m_G} = {\hat x}_{m_G} e^{\zeta({\hat x}_{m_G})} ,
\ee
which inserted into \eqref{rho_NL} gives
\be \label{rho_nl}
\overline{\delta\rho}(\hat{x}) = \delta\rho_{0_G} \left[ \frac{\sin\hat{x}}{\hat{x}} + \frac{27}{2}\frac{\delta\rho_{0_G}}{x^2_{m_G}}  \left( {\cal F}_1({\hat x}) +  \frac{3}{5}f_\NL{\cal F}_2({\hat x}) \right)   \right] 
\exp{ \left[ - \left( \frac{9}{4}\frac{\delta\rho_{0_G}}{x^2_{m_G}} \right)^3 \frac{1+\frac{3}{5}f_\NL}{2\pi A} \right] }.
 \ee
We see that for $A=0$ the perturbation vanishes ($\delta\rho({\hat x})=0$). This can be renormalized 
 with respect the central value as
 \be\label{rho_nl2}
 \overline{\delta\rho}(\hat{x}) = \delta\rho_0 \left[ \frac{ \displaystyle{ \frac{\sin\hat{x}}{\hat{x}} + 
 \frac{\delta\rho_{0_G}}{x^2_{m_G}}  {\cal F}(\hat x) } }
{ \displaystyle{ 1 + \frac{\delta\rho_{0_G}}{x^2_{m_G}} {\cal F}(0)  } } \right],
 \ee
 where  
 \be \label{F_x}
 {\cal F}(x) \equiv \frac{27}{2} \left[ {\cal F}_1(\hat x) + \frac{3}{5}f_\NL {\cal F}_2(\hat x) \right]
 \quad \quad \textrm{and} \quad \quad 
{\cal F}(0) =  \frac{27}{2} \left( 1 + \frac{3}{5}f_\NL  \right) \,.
 \ee
 Finally, the peak amplitude $\delta\rho_0$ of the average energy density profile is related to the amplitude of the peak in the Gaussian approximation 
 $\delta\rho_{0_G}$ as
 \be\label{rho_0}
 \delta\rho_0 =  \delta\rho_{0_G} \left[ 1 + \frac{\delta\rho_{0_G}}{x^2_{m_G}} {\cal F}(0) \right]  
 \exp{ \left[ - \left( \frac{3}{4} \frac{\delta\rho_{0_G}}{x^2_{m_G}} \right)^3 \frac{{\cal F}(0)}{\pi A} \right] } \,.
 \ee
Apart from the exponential correction, which we will see later at the end of Section \ref{Numerical results} that can be usually neglected, the profile 
given by \eqref{rho_nl2} is a second-order expansion in terms of the Gaussian amplitude of the peak, consistently with the second order approach 
we are following. 

 \begin{figure*}
\vspace{-1.5cm}
\includegraphics[width=0.55\textwidth]{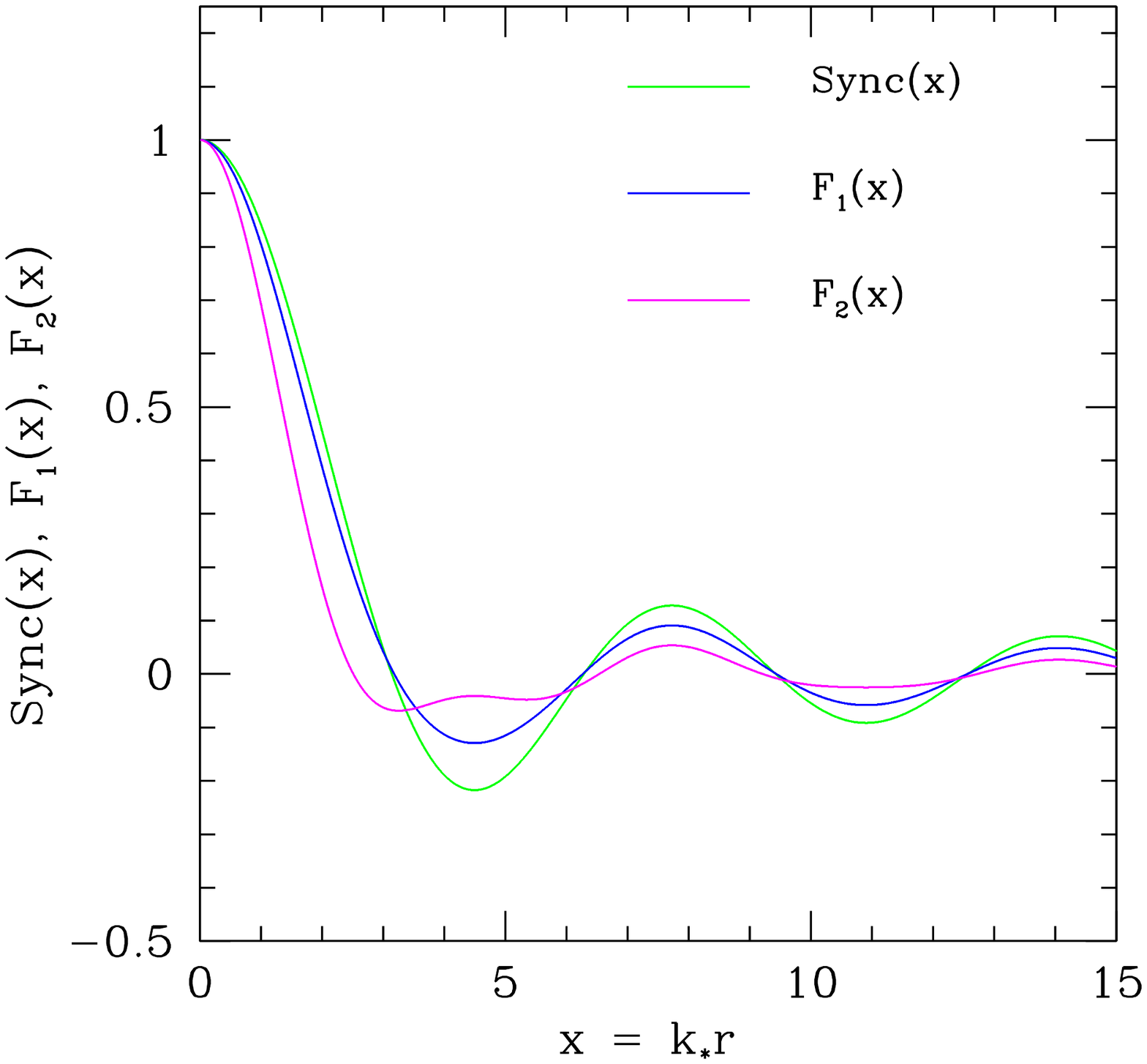} 
  \vspace{-2.5cm}
  \caption{This figure shows the behaviour of the three different components of the shape given by \eqref{rho_nl3} as function of $x=k_*r$: the linear 
  component given by ${\rm Sync}(x)=\sin{x}/x$, the non-linear component ${\cal F}_1(x)$ and the non-Gaussian component ${\cal F}_2(x)$. This allows 
  to appreciate the different steepness of the components of the final shape of the energy density, combined together for different values of $f_\NL$. }
  \label{sync}
\end{figure*}

 We are now going to assume $x \simeq \hat{x}$, neglecting the exponential term, because there is no an analytic form of $\zeta(\hat x)$ corresponding 
 to \eqref{rho_nl2}, necessary to calculate precisely the value of $\hat{x}_m$ and the perturbation of the velocity field given by \eqref{delta_U}.  
 The exponential in \eqref{rho_0} is only a numerical coefficient that is not going to change the profile of \eqref{rho_nl2}, changing only the relative value 
 of $\delta\rho_{0_G}$ with respect the physical value of the peak given by $\delta\rho_0$. The value of the threshold $\delta_c$ and the corresponding 
 critical peak amplitude $\delta\rho_{0c}$ are therefore independent from the value of $A$. To simplify the treatment we are therefore going to neglect this 
 exponential term, keeping in mind that the numerical values of $\delta\rho_{0_G}$ obtained in the next section should be in general associated to 
$\delta\rho_{0_G}e^ {-\mathcal{S}(A)}$, where
\[ \mathcal{S}(A) = \left( \frac{3}{4} \frac{\delta\rho_{0_G}}{x^2_{m_G}} \right)^3 \frac{{\cal F}(0)}{\pi A} \,. \]
According to this we simplify \eqref{rho_nl2} as
 \be\label{rho_nl3}
  \overline{\delta\rho} (x) = \delta\rho_{0_G} \left[ \frac{\sin{x}}{x} + \frac{\delta\rho_{0_G}}{x^2_{m_G}} {\cal F}(x) \right]. 
 \ee
The function ${\cal F}(x)$ is a second-order correction to the profile, measured in powers of  $\delta\rho_{0_G}$, with respect to the linear Gaussian approximation
where ${\cal F}(x)=0$. This is the final form of the profile that will be used to compute numerically in the next section the corresponding value of the threshold $\delta_c$ 
for different values of $f_\NL$. This approximation is consistent with the second-order expansion we have used here to derive the energy density profile. However one 
should remember that, because the threshold of PBH formation is non-linear, in principle all the non-linear components of the curvature perturbation should be taken 
into account. The aim of this calculation is to check if the amplitude of the modification given by the three-point correlation function truncating \eqref{final} at the third 
order, including also a possible non-Gaussian component of the power spectrum, is small. Only in this case our approach would be consistent.

The input parameter measuring the amplitude of the perturbation is given by $\delta\rho_0$, with the corresponding Gaussian value computed with \eqref{rho_0}.
The shape of the energy density profile given by \eqref{rho_nl3} is characterized by three different functions: ${\rm sync}(x)=\sin x/x$, ${\cal F}_1(x)$, ${\cal F}_2(x)$, 
combined together with different coefficients to determine the final shape. In Figure \ref{sync} these functions are plotted against $x=k_*r$, showing that ${\cal F}_2(x)$ 
is a bit steeper than ${\cal F}_1(x)$ which is itself slightly steeper than ${\rm sync}(x)$. Depending on the sign of $f_\NL$, these three functions will combine in different 
ways and the final \emph{non-linear shape} given by \eqref{rho_nl3} would be steeper or shallower with respect to the \emph{Gaussian shape} which is described simply 
by the  sync function. We will see later in Section \ref{Numerical results} how the threshold $\delta_c$ for PBH formation is changing with respect to the linear Gaussian 
case, varying also the value of $f_\NL$.   

 An analogous calculation gives the profile of the velocity field: inserting \eqref{rho_nl3}  into \eqref{delta_U} and  assuming $aHr_m = 1$, we get
 \be
 \delta U(x) = - \frac{1}{1+w} \frac{\delta\rho_{0_G}}{x^3} \left[ {\cal G}_0(x) + \frac{\delta\rho_{0_G}}{x^2_{m_G}} {\cal G}(x) \right],
 \ee
 where the functions ${\cal G}_0(x),{\cal G}(x)$ are defined as
 \[
{\cal G}_0(x) \equiv \int_0^{x} \frac{\sin{x}}{x} x^2\d x =  \sin{x} - x\cos{x} \,,
{\cal G}(x) \equiv \int_0^{x} {\cal F}(x)x^2\d x = \frac{27}{2} \left[  {\cal G}_1(x) + \frac{3}{5}f_\NL {\cal G}_2(x)  \right].
 \]
The integrals of the functions ${\cal F}_{1,2}(x)$ can be computed analytically
 \begin{eqnarray}
 {\cal G}_1(x) & \equiv & \int_0^{x} {\cal F}_1(x)x^2\d x = \frac{1}{3} \left[ 2{\cal G}_0(x) + \frac{1}{8} \left( 5x + \frac{\cos{2x}-1}{x} - \frac{3}{2}\sin{2x} \right) \right],\nonumber \\
 {\cal G}_2(x) & \equiv & \int_0^{x} {\cal F}_2(x)x^2\d x =  \frac{1}{3} \left[ {\cal G}_0(x) - \sin{2x} - \frac{\cos{2x}-1}{x} \right], \nonumber
 \end{eqnarray}
where ${\cal G}_0(0) = {\cal G}_1(0) = {\cal G}_2(0) = 0$ as one would expect consistently with the boundary condition of the velocity at the centre ($U(0) = 0$). 
We notice that the function ${\cal G}(x)$ is formally modifying the profile of the velocity field with respect to the linear Gaussian case given by ${\cal G}_0(x)$, as 
the function  ${\cal F}(x)$ is doing for the energy density profile given by $\sin{x}/x$. 

For the numerical implementation of this perturbation, we need to compute the value of $x_ m$ in terms of the initial input parameters, that is  the amplitude measured 
by the central peak $\delta\rho_0$, the peak of the power spectrum $A$ and the non-Gaussian component of the power spectrum measured by $f_\NL$. The integral 
relation for $x_m$ as follows from \eqref{rho_nl3} and \eqref{r_m} is  explicitly written as 
 \be\label{x_m}
  \left( x_m^2 - 1 \right) \sin{x_m}  + x_m\cos{x_m} + \frac{\delta\rho_{0_G}}{x^2_{m_G}} 
  \left[ x_m^3 {\cal F}(x_m) - {\cal G}(x_m) \right]  = 0\,, 
  \ee
 and needs to be solved numerically. When  ${\cal F}(x)=0$, which implies that ${\cal G}(x)=0$,  one gets $x_{m_G}\simeq2.74$ consistently with \cite{musco}. Finally 
 we are now able to calculate the averaged amplitude $\delta_m$ from the input value of the central energy density peak $\delta\rho_0$ inserting \eqref{rho_nl3} into 
 \eqref{delta_m} for $x=x_m$.

 \subsection{The density profile from the averaged curvature profile $\bar\zeta$}
 \label{Averaged curvature}
 \noindent
 In the following we are going to derive the energy density profile corresponding to the mean curvature profile $\bar \zeta$ obtained from the peaked power spectrum 
 when peak theory is applied to the Gaussian variable $\zeta$ instead of the standard approach using the energy density $\delta\rho$. The two approaches in general 
 are not equivalent because of the non-linear relation of expression \eqref{rel}: even though peaks in $\zeta$ correspond to peaks in $\delta\rho$ if they are steep enough \cite{haradath,ng3},  the energy density profile obtained with this from $\bar \zeta$ does not correspond to the mean profile of the energy density. The aim here is to 
 compare in the next section the threshold of this profile with the one obtained earlier in \eqref{rho_nl3}. The mean curvature profile $\bar \zeta$ corresponding to a peaked 
 power spectrum is \cite{haradath}
 \be\label{zeta_peak}
\bar{\zeta}(\hat r) = \zeta_0  \frac{\sin{k_*\hat r}}{k_*\hat r}
 \ee
and plugged into \eqref{deltarho_zeta} gives
\be\label{delta_rho_zpeak}
\delta\rho (\hat{x},t) = \frac{4}{9} \left( \frac{k_*}{aH} \right)^2 \left[\overline\zeta(\hat x) - 
\frac{1}{2} \left( \frac{ \zeta_0 \cos{\hat x} - \overline\zeta(\hat x) } {{\hat x}} \right)^2 \right] e^{-2\overline\zeta(\hat x)} \,,
\ee
where $\hat x \equiv k_* \hat x$. The overdensity at the center turns out then to be 
\be
\delta\rho_0 (0,t) = \frac{4}{9} \left( \frac{k_*}{aH} \right)^2 \zeta_0 e^{-2\zeta_0},
\ee
which allows to renormalize \eqref{delta_rho_zpeak} as
\be\label{delta_rho_zpeak2}
\delta\rho (\hat{x},t) = \delta\rho (0,t) \left[ \frac{\sin{\hat x}}{\hat x} - 
\frac{1}{2} \left( \frac{ {\hat x}\cos{\hat x} - \sin{\hat x}}{{\hat x}^2} \right)^2 \right] 
\exp{\left[ - 2\zeta_0  \left( \frac{\sin{\hat x}}{\hat x} -1 \right)\right] } \,.
\ee
We can then calculate  the scale ${\hat x}_m$ of the perturbation by solving equation \eqref{eq_rm}, which is explicitly written as 
\be \label{hx_m}
({\hat x}_m^2-1)\sin{{\hat x}_m} + {\hat x}_m\cos{{\hat x}_m} = 0\,,
\ee 
analogous to \eqref{x_m} when ${\cal F}(x)=0$ and 
its solution is ${\hat x}_m\simeq2.74$. Because the horizon crossing is calculated in real space when $aHr_m=1= aH{\hat r}_m e^{\zeta({\hat r}_m)}$,  
it is necessary to renormalize the central peak of the energy density with respect to $x_m = {\hat x}_m e^{\zeta({\hat x}_m)}$, that is
\be \label{delta_rho0-zeta}
\delta\rho (0,t) =  \frac{4}{9} \left( \frac{1}{aHr_m} \right)^2 x_m^2 \zeta_0 e^{-2\zeta_0} 
\quad\quad \Rightarrow \quad\quad 
\delta\rho_0 = \frac{4}{9} x_m^2 \zeta_0 e^{-2\zeta_0}  \,.
\ee
Using now the expression for $\delta_m$ given by \eqref{delta_m},  we find that in terms of $\zeta({\hat x}_m)$ 
\be
\delta_m = - \frac{2}{3} {\hat x}_m \zeta'({\hat x}_m)  \left( 2 + {\hat x}_m \zeta'({\hat x}_m) \right) ,
\ee
which combined with \eqref{zeta_peak} and \eqref{hx_m} leads to
\be
\zeta_0 = \frac{1-\hat{x}^2_m}{\hat{x}^2_m \cos{\hat{x}_m}} \left[ 1 - \sqrt{1 - \frac{3}{2}\delta_m} \,\right] 
\simeq 0.94 \left[ 1 - \sqrt{1 - \frac{3}{2}\delta_m} \, \right] \,.
\ee
Replacing this into \eqref{delta_rho0-zeta}, one can calculate the peak amplitude of the energy density from the averaged 
perturbation amplitude $\delta_m$.

\section{Numerical results}
\label{Numerical results}\noindent
The averaged profiles of the density, velocity and curvature profiles analyzed in the previous section have been implemented as initial conditions, 
using the gradient expansion approach described in Section \ref{Initial Conditions} to calculate the corresponding threshold of PBH formation with
the same code used in \cite{Musco:2004ak,Polnarev:2006aa,Musco:2008hv,Musco:2012au,musco,ng4}. This has been fully described previously 
and therefore we give only a very brief outline of it here. 
It is an explicit Lagrangian hydrodynamics code with the grid designed for calculations in an  
expanding cosmological background. The basic grid uses logarithmic spacing in a mass-type comoving coordinate, allowing it to reach out to very 
large radii while giving finer resolution at small radii necessary to have a good resolution of the  initial perturbation. The initial data are specified 
on a space-like slice at constant initial cosmic time $t_i$ defined as $a(t_i)r_m = 10/H$, ($\epsilon = 10^{-1}$), while the outer edge of the grid has 
been placed at $90 R_m$, to ensure that there is no causal contact between it and the perturbed region during the time of the calculations. The initial 
data are  evolved using the Misner-Sharp-Hernandez equations so as to generate  a second set of initial data on an initial null slice which are then 
evolved using the Hernandez-Misner equations. 
During the evolution, the grid is modified with an adaptive mesh refinement scheme (AMR), built 
on top of the initial logarithmic grid, to provide sufficient resolution to follow black hole formation down to extremely small values of ($\delta-\delta_c$).
The code has a long history and has been carefully tested in its various forms. Numerically is a second order scheme, using double precision 
(16 digits after the coma) keeping the numerical error less than $10^{-4}$ . Further information about tests of the code, including a convergence test, 
could be found in the appendix B of \cite{Musco:2008hv}.

\begin{figure*}[t!]
\vspace{-2.0cm}
 \includegraphics[width=0.49\textwidth]{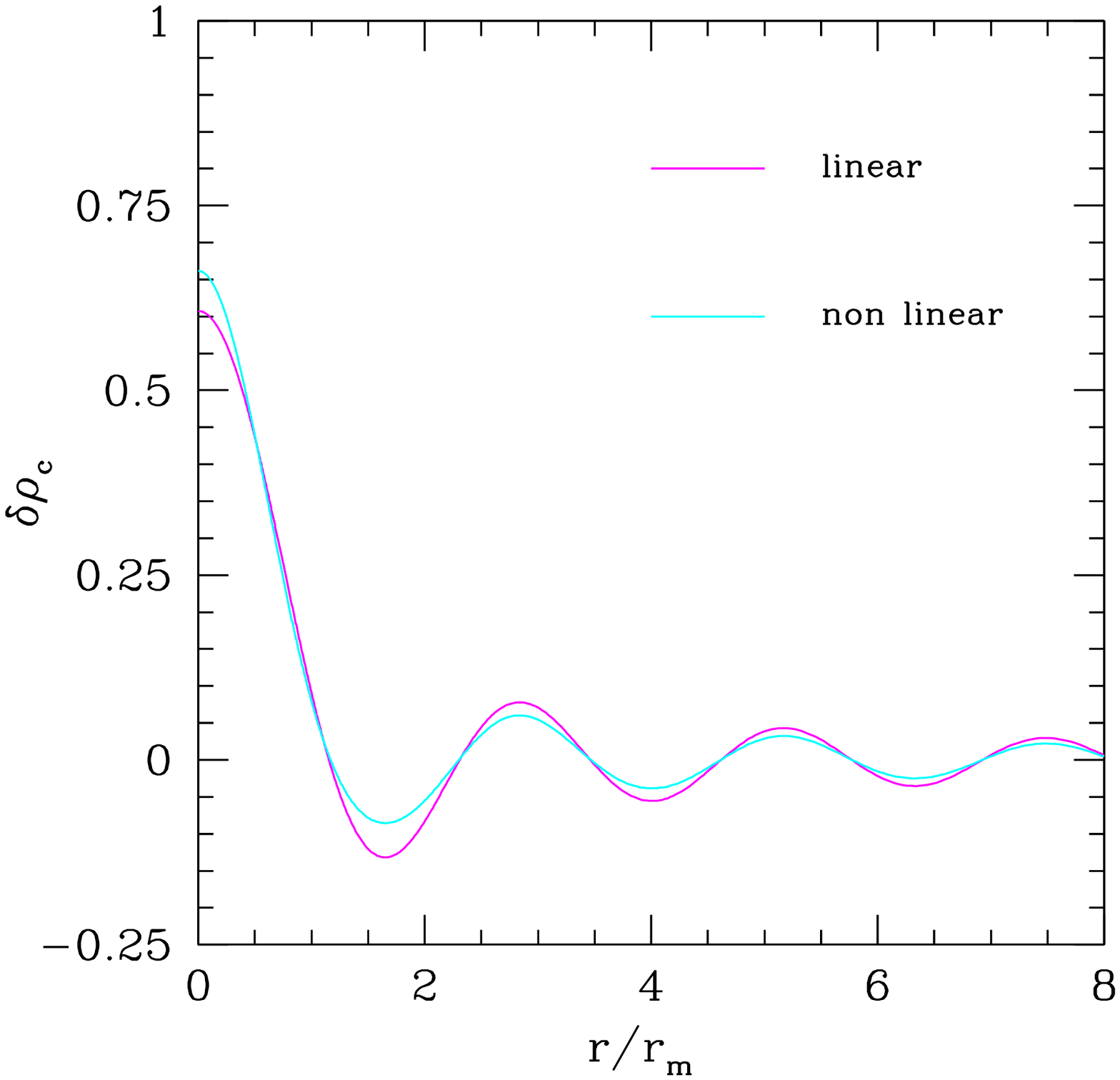} 
  \includegraphics[width=0.49\textwidth]{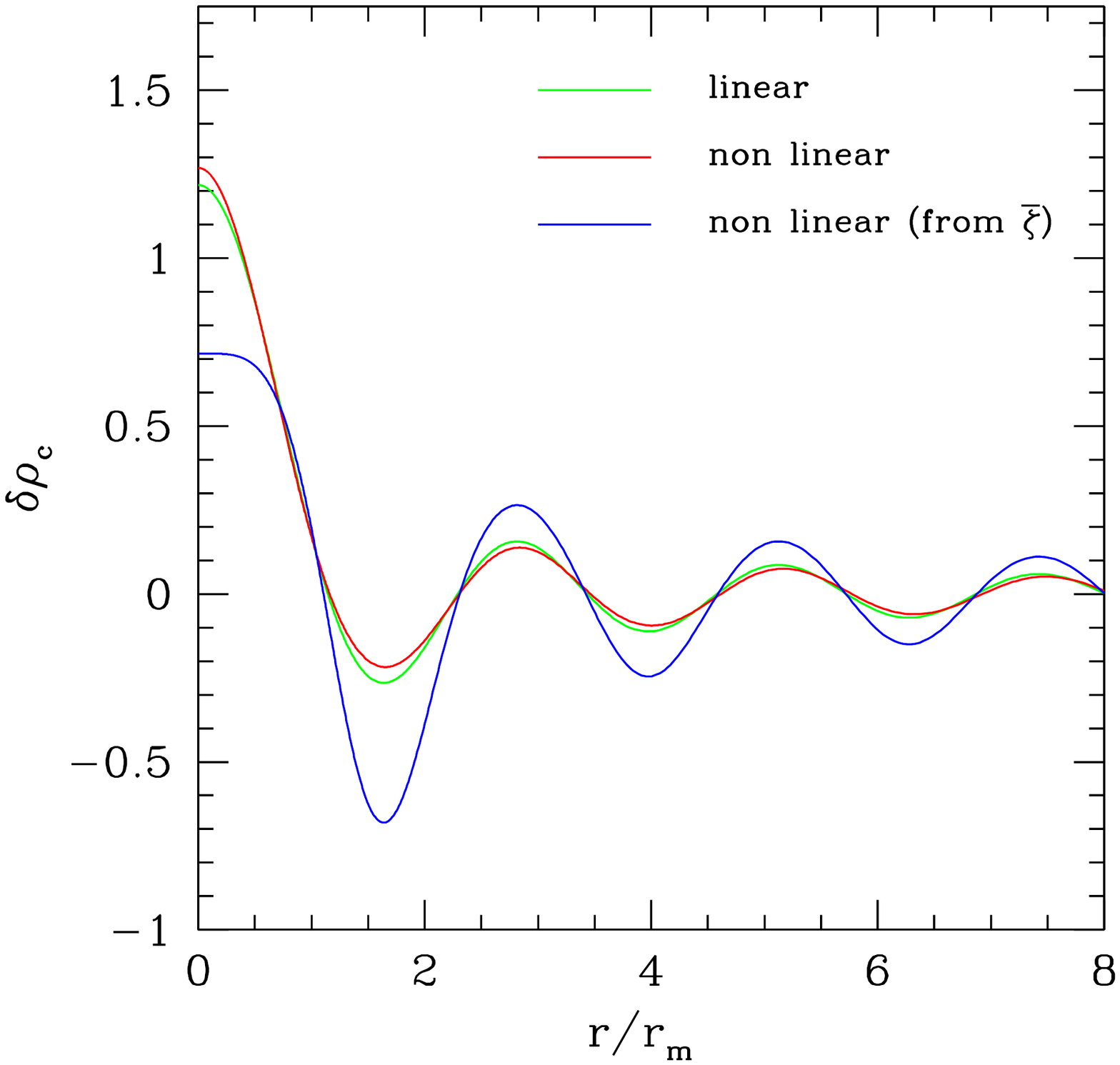} 
  \vspace{-2.5cm}
  \caption{ The left plot shows the two different components of the average shape when $\zeta$ is Gaussian ($f_\NL=0$), respectively the linear one in 
  magenta and the non linear one in cyan. In the right plot we combine them together obtaining the full average non linear shape of the density
  $\overline{\delta\rho}$ (red line) that can be compared with the average Gaussian density profile obtained with the linear approximation (green line). 
  The blu line instead shows the corresponding non Gaussian density profile computed directly from $\overline{\zeta}$ using the non linear relation 
  between the energy density and the curvature profile. All these quantities are plotted against $r/r_m$. }
  \label{critical_profiles}
\end{figure*}

We are now going to analyze the critical average profiles given by \eqref{rho_nl3} showing explicitly the different components that gives rise to the final profile,
when $\zeta$ is a Gaussian random variable ($f_\NL=0$) and when a non-Gaussian contribution to the field is also taken into account ($f_\NL\neq0$), using 
both positive and negative values of the non-Gaussian parameter. One can write explicitly the different components as
\bea
& & \left. \overline{\delta\rho} \right\vert_{\rm linear}  \ \ \ \ \ \,  = \  \delta\rho_{0_G} \left( \frac{\sin{x}}{x} \right) \,, \label{linear_comp}\\ \nonumber\\
& & \left. \overline{\delta\rho} \right\vert_{\rm non\, lin.}  \ \ \ \, =  \ \frac{27}{2} \left( \frac{\delta\rho_{0_G}}{x_{m_G}} \right)^2 {\cal F}_1(x) \label{nlinear_comp} \,, \\ \nonumber\\
& & \left. \overline{\delta\rho} \right\vert_{\rm non\, Gauss.}  = \  \frac{81}{10} f_\NL \left( \frac{\delta\rho_{0_G}}{x_{m_G}} \right)^2 {\cal F}_2(x) \label{nGaussian_comp}\,, \\ \nonumber\\
& & \left. \overline{\delta\rho} \right\vert_{\rm total} \ \ \ \ \ \ \, = \ \delta\rho_{0_G} \left[ \frac{\sin{x}}{x} + \frac{27}{2} \frac{\delta\rho_{0_G}}{x^2_{m_G}}  \label{total_profile}
\left( {\cal F}_1(x) + \frac{3}{5} f_\NL {\cal F}_2(x) \right) \right] \,.
\eea
and we notice that the non-linear components are one order or magnitude higher in terms of $\delta\rho_{0_G}/\rho_b$, consistent with our perturbative 
approach. Because $27/2x^2_{m_G}\simeq1.80$, if the peak amplitude of the perturbation is small ($\delta\rho_{0_G} \ll 1$), then the non-linear components can be 
neglected and linear theory can be used with good accuracy to calculate the shape of the average density peak, while if $\delta\rho_{0_G} \sim 1$ or larger, as it is 
necessary for PBH formation \cite{musco}, the non-linear components have the same amplitude of the linear one and one should take them into account. 

\begin{figure*}[t!]
\vspace{-1.5cm}
  \includegraphics[width=0.49\textwidth]{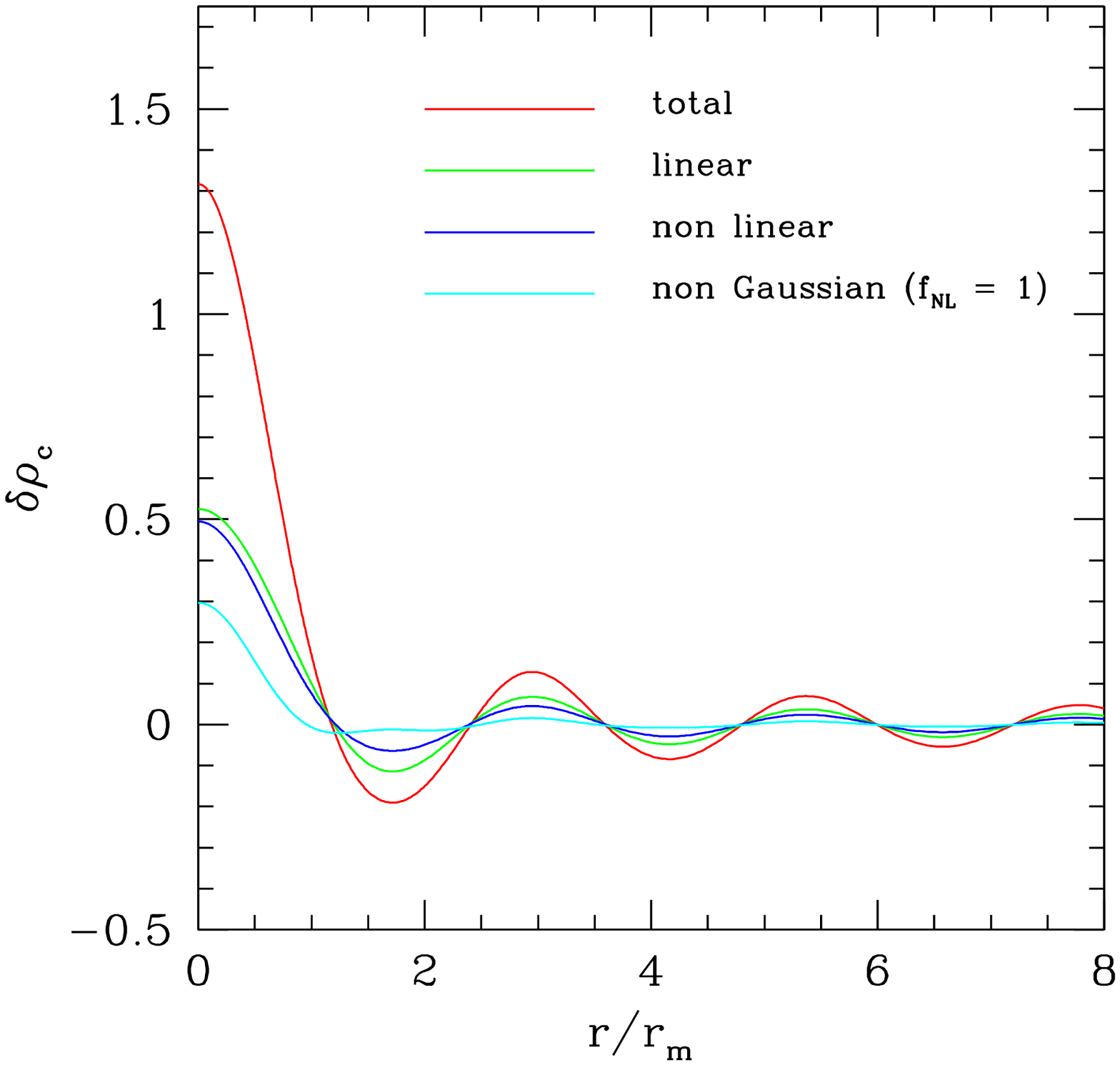} 
   \includegraphics[width=0.49\textwidth]{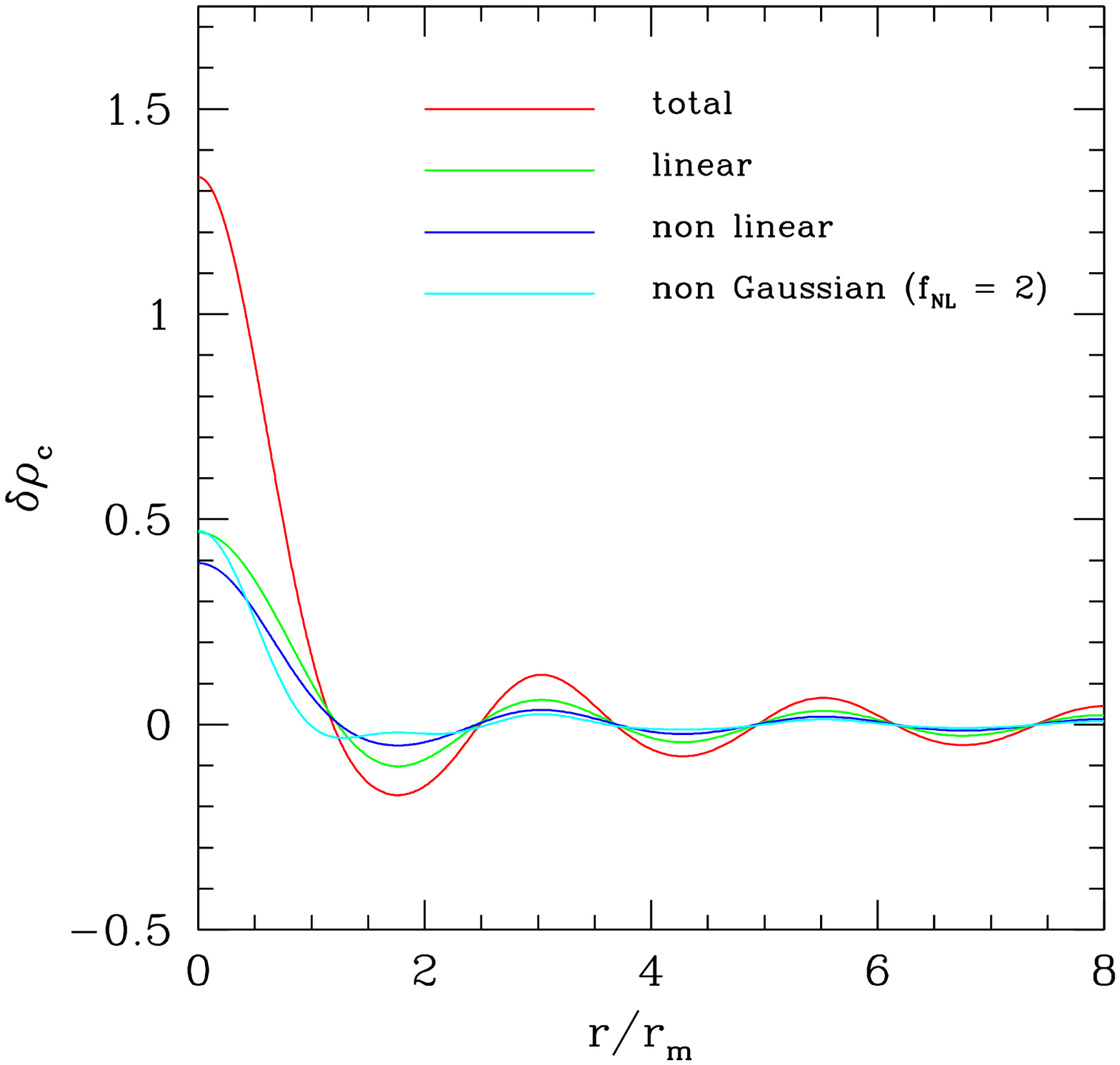}
     \end{figure*}
  \begin{figure*} [t!]
  \vspace{-4.5cm}
   \includegraphics[width=0.49\textwidth]{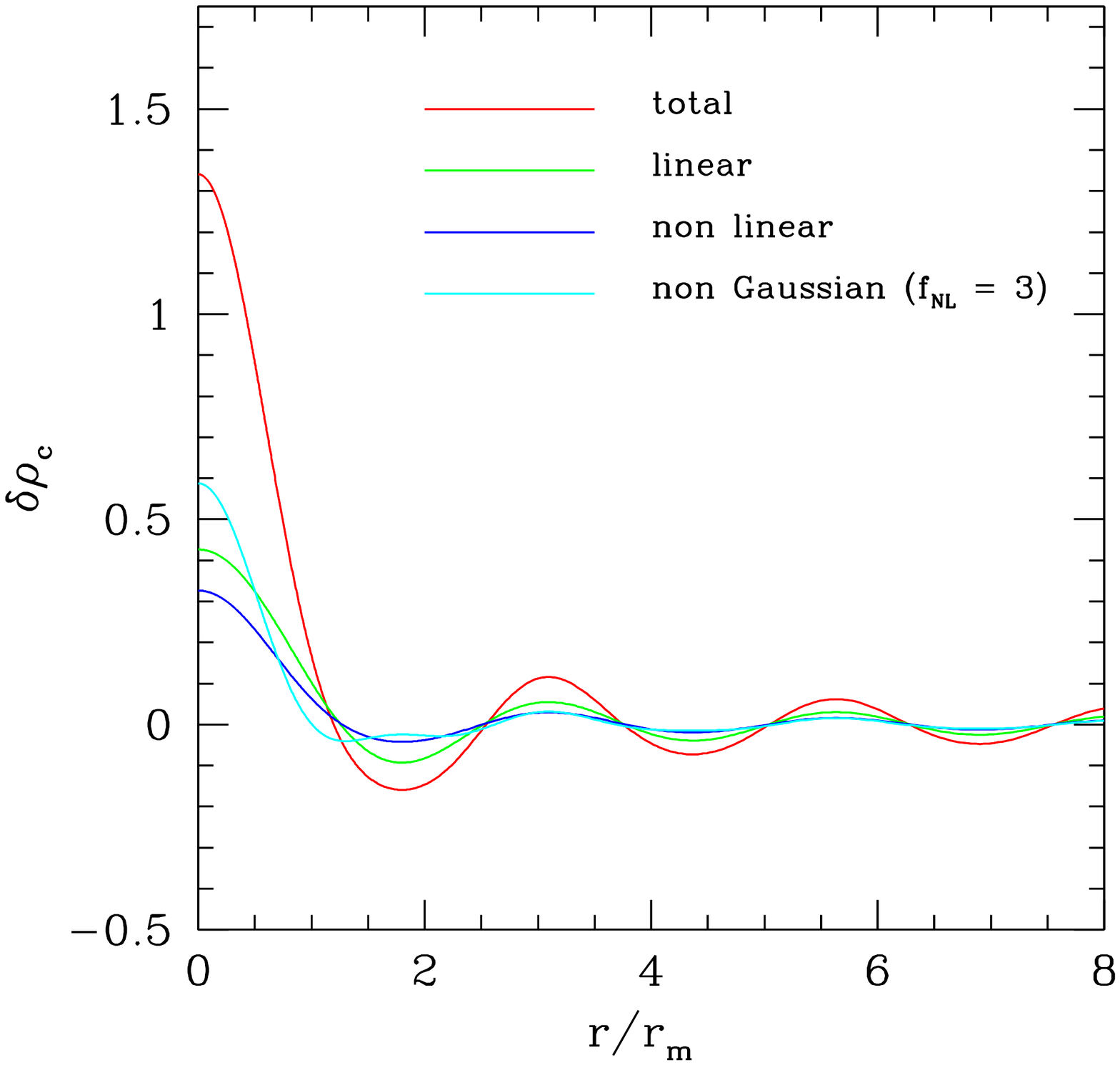} 
   \includegraphics[width=0.49\textwidth]{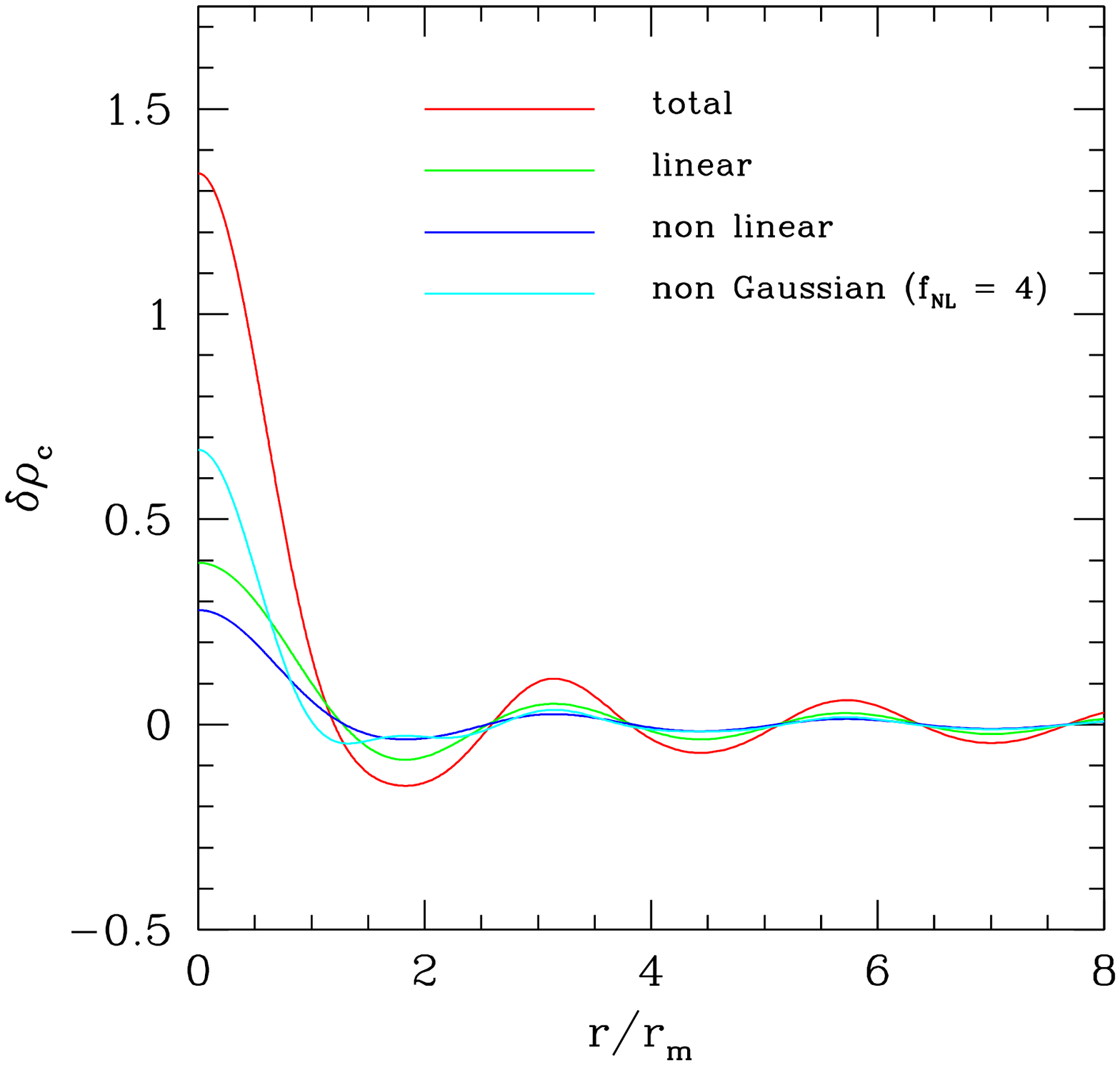}
  \vspace{-2.5cm}
  \caption{The  different components of the critical profile of the energy density for 4 different cases ($f_\NL=1, 2, 3, 4$), all plotted against $r/r_m$. 
  The red line represents the critical profile obtained by the sum of 3 components shown here separately: the linear one (green line), the non-linear one (blue line) and the 
  non-Gaussian one (cyan line). }
  \label{critical_profiles_fnl}
\end{figure*}

In principle this is questioning our second-order expansion approach to compute the threshold for PBH formation $\delta_c$, suggesting that one should compute all the 
higher-order terms of \eqref{final} for an accurate computation, which would be extremely difficult. However, because the shapes of the all three components are similar 
to each other in the range of the overdensity region (see Figure \ref{sync}), it will turn out that the final shape is not very different from the linear one, because is a
combination of three similar shapes. For this reason, the final critical amplitude of the peak is not very different from the one calculated with the linear approximation 
as we can see in the right plot of Figure \ref{critical_profiles} where we are comparing the linear critical average density profile (green line) with the one obtained using 
the non linear approximation (red line) obtained by the combination of the linear and non-linear components, represented separately in the left plot with the magenta and 
cyan lines respectively. As we have argued these two components have a comparable amplitude, but the two final linear and non-linear shapes are very similar, and the 
threshold $\delta_c$ of the non linear case is about $1\%$ smaller in the linear case, while the critical amplitude of the peak is about $4\%$ larger in the non linear case 
with respect the linear one.

\begin{figure*}[t!]
\vspace{-1.5cm}
  \includegraphics[width=0.49\textwidth]{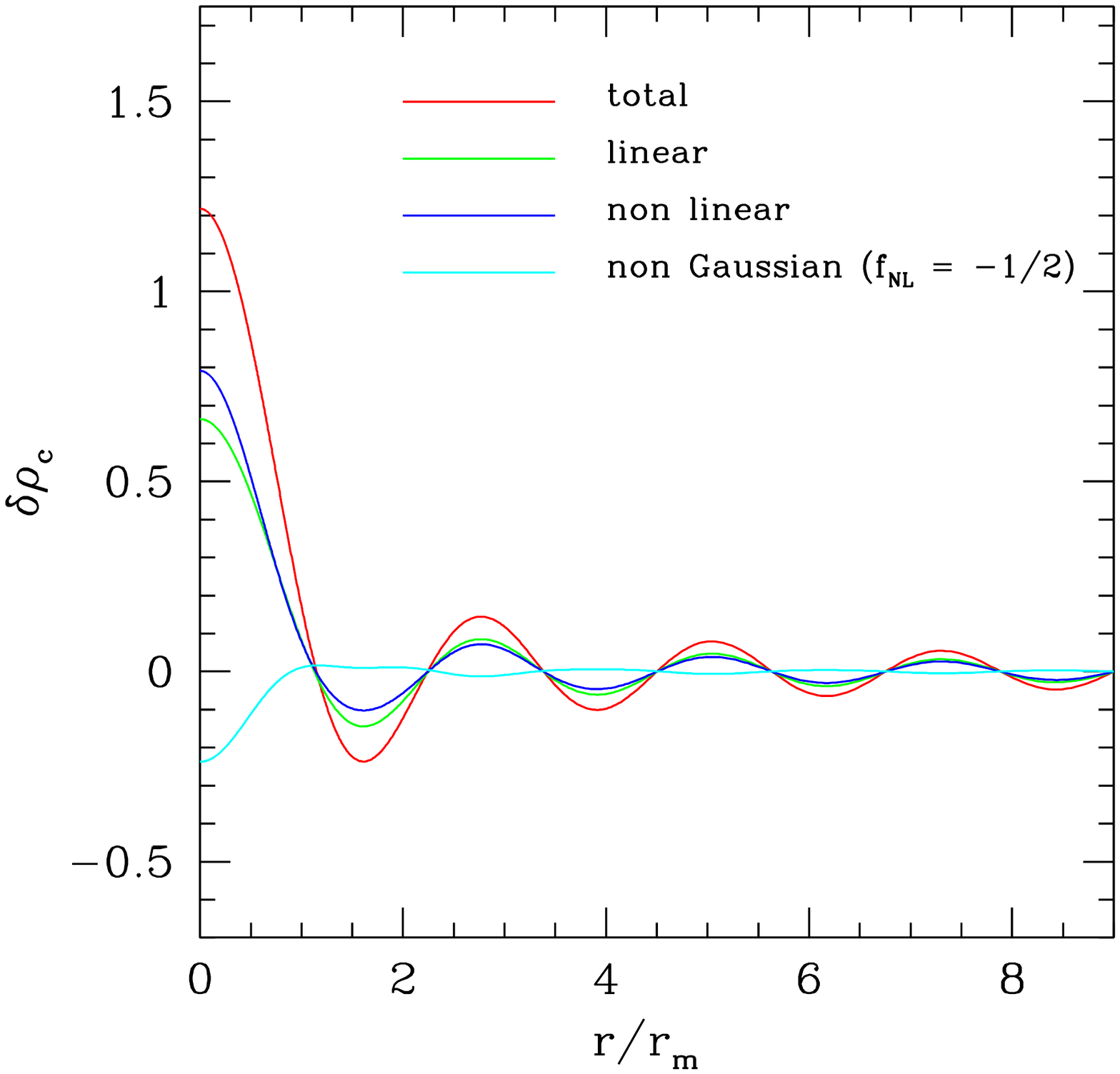} 
   \includegraphics[width=0.49\textwidth]{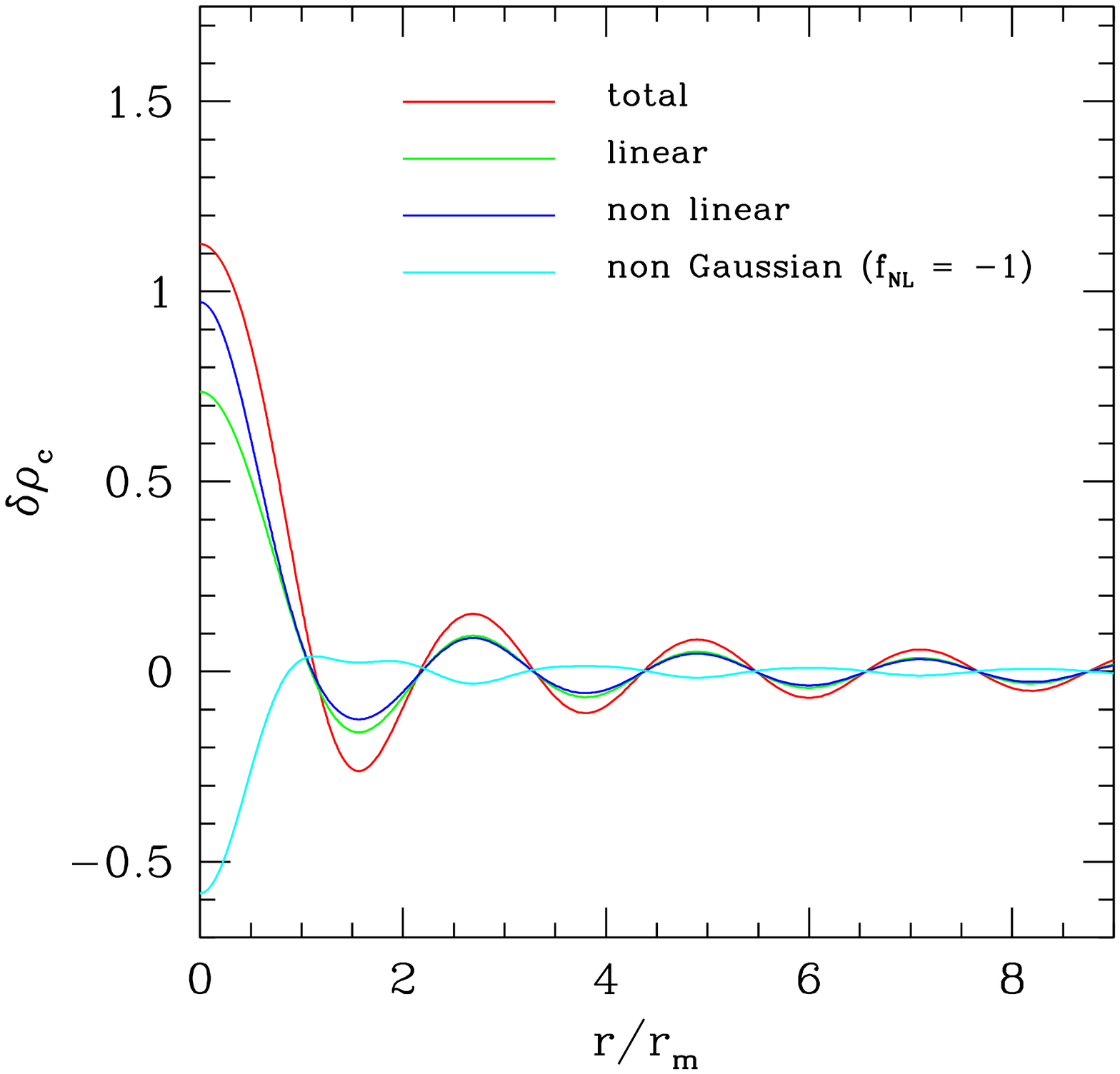}
   \end{figure*}
   \begin{figure*} [t!]
   \vspace{-4.5cm}
   \includegraphics[width=0.49\textwidth]{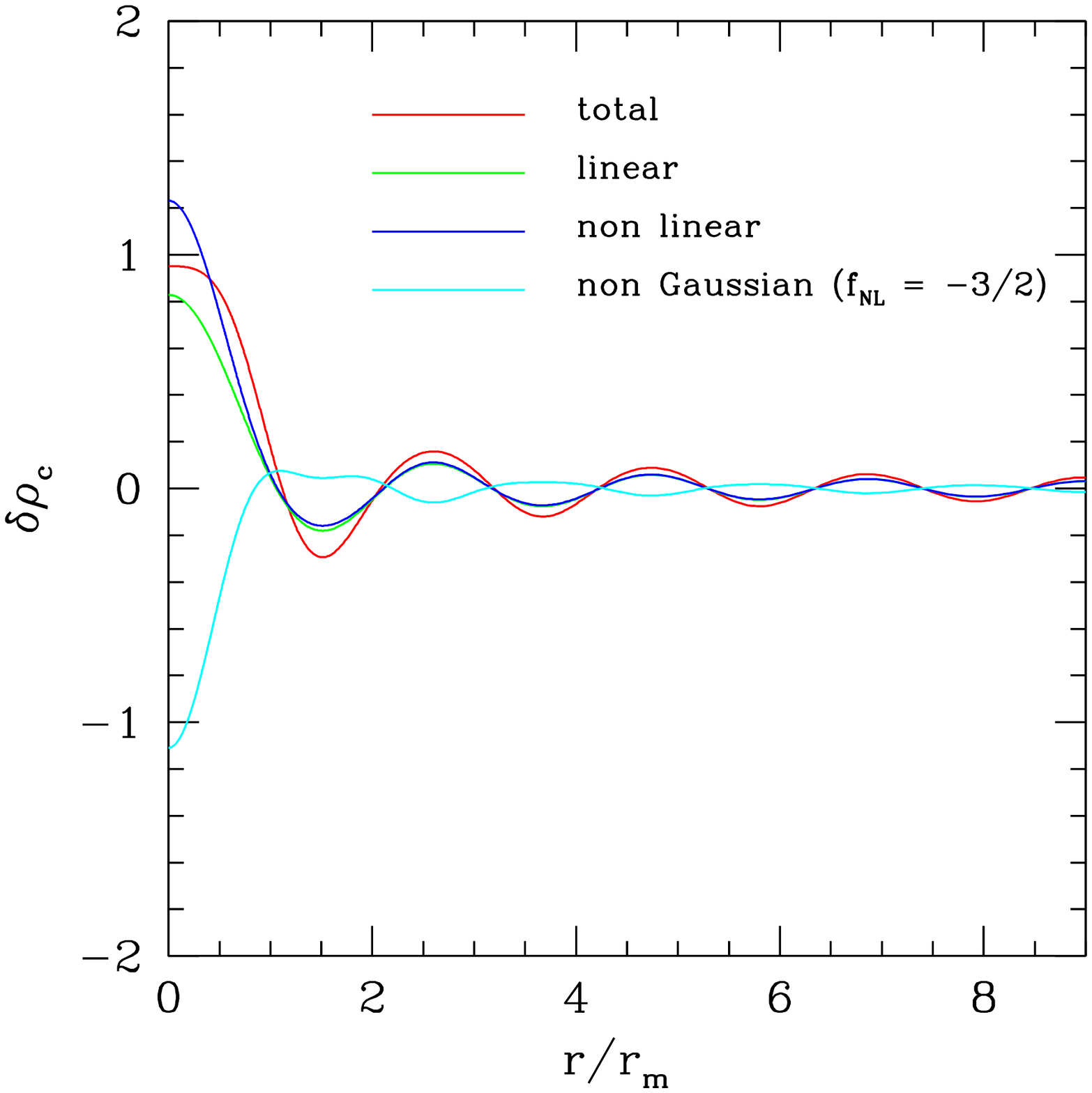} 
   \includegraphics[width=0.49\textwidth]{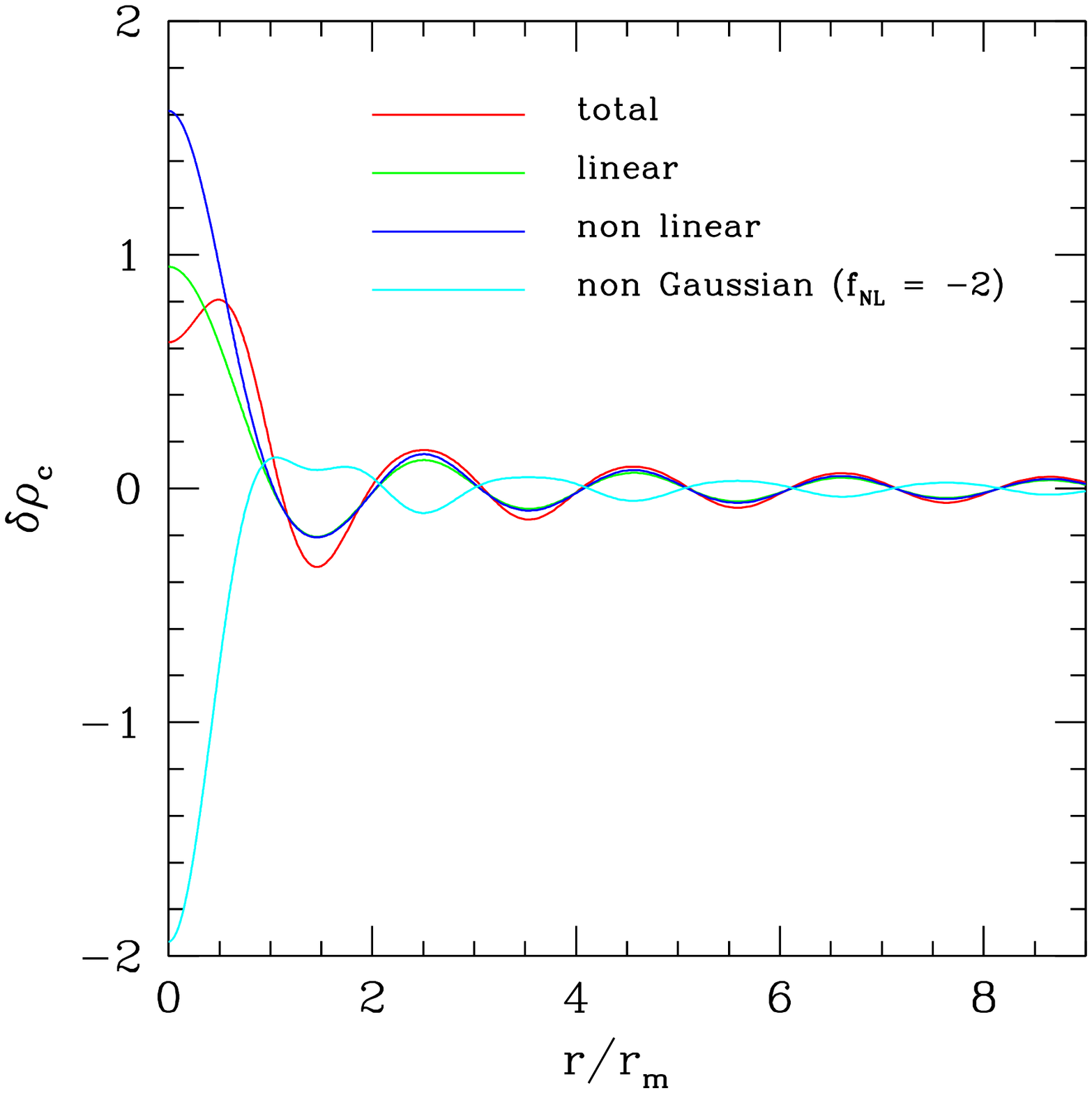}
  \vspace{-2.5cm}
  \caption{The  different components of the critical profile of the energy density for 4 different cases ($f_\NL=-1/2, -1, -3/2, -2$), all plotted against $r/r_m$. 
  The red line represents the critical profile obtained by the sum of 3 components shown here separately: the linear one (green line), the non-linear one (blue line) and the 
  non-Gaussian one (cyan line). }
  \label{critical_profiles_fnl2}
\end{figure*}

The blue line in the right plot of Figure \ref{critical_profiles} represents the density profile obtained with the alternative approach of using the averaged profile $\overline{\zeta}$
discussed in Section \ref{Averaged curvature} to compute the profile of the density contrast, which is very different from the one we have obtained
with our perturbative approach. In this case the critical amplitude of the peak is significantly smaller ($\delta\rho_0 = 0.716$ in the non-linear case against 
$\delta\rho_0$ = 1.218 of the linear one) with a difference larger than $40\%$. Because the relation to compute the energy density profile from the curvature profile 
given by (2.4) is non-linear, the mean profile of $\zeta$ does not give the corresponding mean profile of the energy density. Until it will not be clear how this expression 
should be modified, it is not clear if the application of peak theory in $\zeta$ could be used to compute consistently the correct value of the threshold associated to the 
mean energy density profile that one needs to calculate the abundance of PBHs, as it has been done in \cite{haradath,b4}.

In Figure \ref{critical_profiles_fnl} we analyze the critical shape of the density contrast for positive values of $f_\NL$ between $1$ and $4$: when $f_\NL = 1, 2$  
(top plots), the three components (linear, non-linear and non-Gaussian) have a similar amplitude while for $f_\NL = 2, 4$ (bottom plot) the amplitude of the non-Gaussian 
component is larger than the other two and becomes progressively more and more dominant for increasing values of  $f_\NL$. In Figure \ref{critical_profiles_fnl2} the same 
analysis is done for negative values of $f_\NL$ between $-1/2$ and $-2$, where now the non-Gaussian component is negative. 

\begin{figure*}
\vspace{-1.5cm}
  \includegraphics[width=0.49\textwidth]{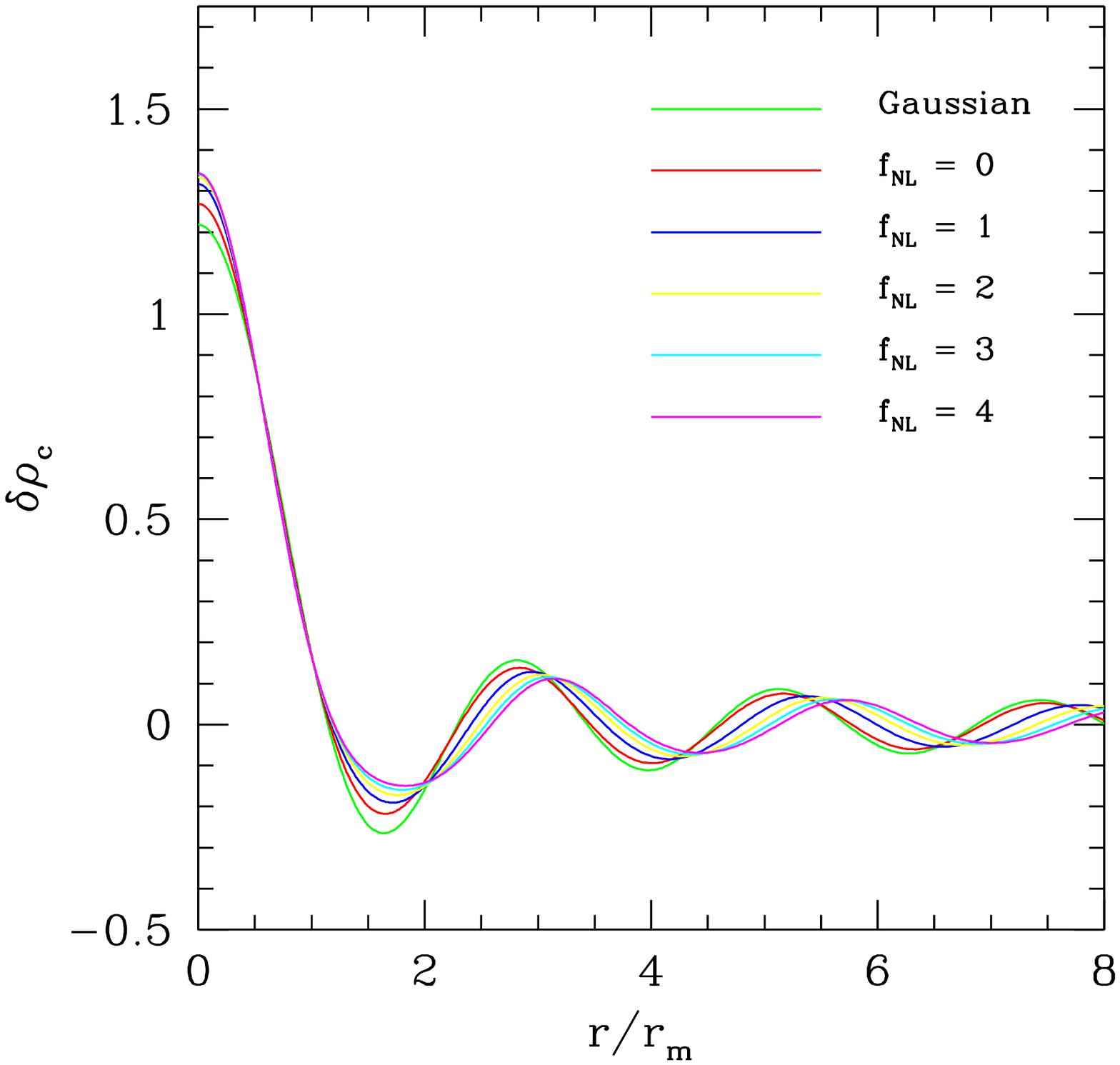} 
  \includegraphics[width=0.49\textwidth]{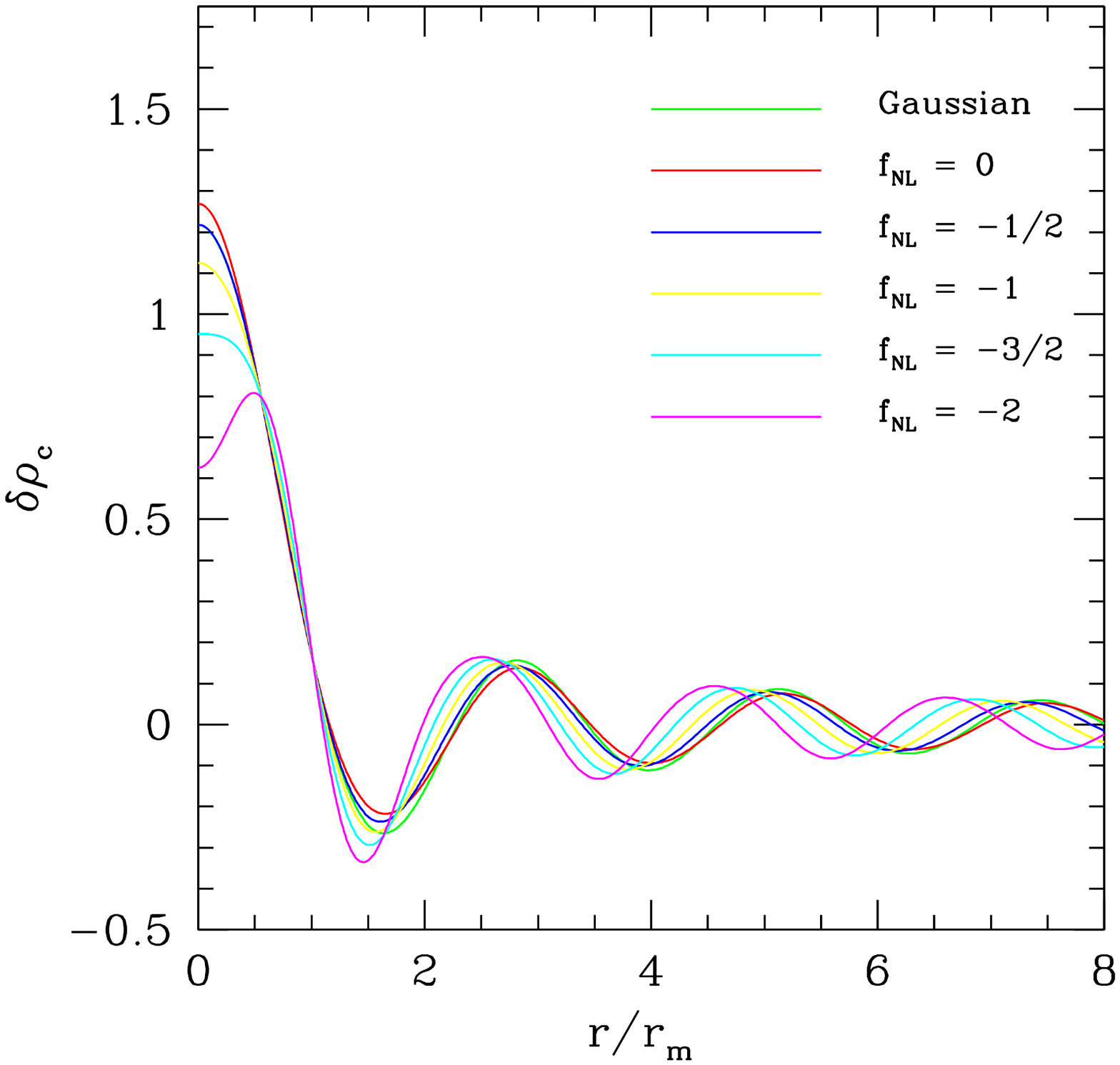} 
  \vspace{-2.5cm}
  \caption{This figure is summarizing the results of all the previous figures comparing the critical shape of the energy density obtained with the linear approximation 
  (green line) with the critical shapes obtained including the non-linear contribution coming from the 3-point correlation function plus the non-Gaussian component of 
  the peaked power spectrum for positive values of $f_\NL$ in the left plot and for negative vales in the right one. In both plots all the density profiles are plotted against 
  $r/r_m$. }
   \label{critical_profiles2}
\end{figure*}

All these pictures show that one can interpret the final critical peak as the combination of 3 different peaks with a similar length-scale, and in the case of positive value of 
$f_\NL$ the three components are all positive, with the two non linear components being slightly steeper, as one can see from Figure \ref{sync}. Therefore the critical 
amplitude of the final peak is progressively increasing with increasing values of $f_\NL$ with respect to the critical shape obtained using a linear approximation as one 
can see from the left plot of Figure \ref{critical_profiles2} where all the critical shapes for $f_\NL\geq0$ are plotted as function of $r/r_m$. For negative values of $f_\NL$ 
instead the non-Gaussian component is a negative peak, with opposite sign with respect to  the other two components. 

In the right plot of Figure \ref{critical_profiles2} we are plotting all the critical shapes for 
$f_\NL\leq0$, and we see that the non-Gaussian component is basically compensating the non-linear component for $f_\NL-1/2$, giving rise to almost the same profile of 
the linear case (in the overdensity region the green and the blue line are indistinguishable), while for more negative values of $f_\NL$ the negative non-Gaussian component 
becomes more and more important, and an off-centered peak arises when $f_\NL=-2$. This is in the limit of what is possible to be  studied with our perturbative approach 
because the value of the coefficient $\delta\rho_\G$ is obtained from equation \eqref{rho_0} neglecting the exponential term (see later for a comment about this approximation) 
which gives a critical value of $f_\NL$ 
\be
f_{\NL_c} = -\frac{5}{3} \left( 1 + \frac{x^2_{m_G}}{54\delta\rho_0} \right),
\ee
beyond which the value of $\delta\rho_\G$ becomes an imaginary number and our perturbative approach breaks down (note that replacing the values from Table \ref{table} 
one obtains indeed $f_{\NL_c} \simeq - 2.04$). This is physically saying that for large enough negative values of $f_\NL$, the non-Gaussian component will dominate and a 
negative peak will arise in the center, which cannot be treated consistently with peak theory because the critical amplitude of the peak is decreasing significantly if the 
perturbation is not anymore centrally peaked, even if the perturbation has the same value of the threshold $\delta_c$. This suggests that a more proper calculation that would 
take into account also the contribution from the higher order correlators in equation \eqref{final} might significantly change the behaviour of the peak.

In Table \ref{table} we have summarized all the numerical values of the main quantities characterizing the critical cases we have studied, divided in three parts. The first part
of the table gives the value of the critical profiles we have shown in Figure \ref{critical_profiles} when $f_\NL=0$, while the second and the third part refer instead to the cases 
of positive and negative values of  $f_\NL$ that we have seen separately in Figure \ref{critical_profiles_fnl} and \ref{critical_profiles_fnl2}, and summarized in the left and right 
plot of Figure \ref{critical_profiles2}. The first two columns of data of the table give the corresponding critical values of the peak amplitude $\delta\rho_c$ and of the threshold 
$\delta_c$ respectively, while the third column gives the corresponding value of $x_m=k_*r_m$. The fourth column gives the ratio between $r_0$, measuring  the edge of the 
overdensity and the typical perturbation scale $r_m$. As seen in \cite{musco}, this is one of the crucial parameters, together with the peak amplitude $\delta\rho$ and the mass 
excess $\delta$ which  characterize the shape. The fifth column gives the amplitude of the corresponding Gaussian peak $\delta\rho_\G$, which is equal to the critical amplitude 
of the peak for the linear case when there are no non-linear corrections to the shape, while for the other cases this value is a coefficient weighting the amplitude of the different 
components of the profile as we have seen in Eqs. \eqref{linear_comp}, \eqref{nlinear_comp} and  \eqref{nGaussian_comp}. Finally the last two columns give the percentage 
fractional correction of the critical value of the peak and of the average threshold with respect to the linear case. 

\begin {table}[t!]
\caption {Values and variations of the basic quantities.} 
\label{table} 
\begin{center}
\begin{large}
\begin{tabular}{ |c c c c c c c c | } 
\hline
\ \ Type \ \ & \ \ \ $\delta\rho_c$ \ \ \ \ & \ \ \  $\delta_c$ \ \ \  & \ \ \ $x_m$ \ \ \ & \ \ $\sfrac{r_0}{r_m}$ \ \ & \ \ $\delta\rho_{0_G}$ \ \ & \ \ $\Delta\delta\rho_0$ \ \ & \ \ $\Delta\delta_c$  \ \ \\ 
\hline\hline
linear & 1.218 & 0.516 & 2.744 & 1.145 & 1.218 & $-$ & $-$  \\
$f_\NL=0$ & 1.269 & 0.511 & 2.722 & 1.160 & 0.607 & 0.042 & 0.010  \\ 
\  $\bar{\zeta}_\NL$ & 0.716 & 0.582 & 2.989 & 1.027 & $-$ & 0.412 & 0.129  \\
\hline
$f_\NL=1$ & 1.317 & 0.507 & 2.620 & 1.178 & 0.507 & 0.081 & 0.017 \\  
$f_\NL=2$ & 1.334 & 0.504 & 2.549 & 1.189 & 0.468 & 0.095 & 0.022 \\
$f_\NL=3$ & 1.341 & 0.503 & 2.497 & 1.197 & 0.427 & 0.101 & 0.025 \\
$f_\NL=4$ & 1.343 & 0.502 & 2.458 & 1.201 & 0.394 & 0.103 & 0.027  \\ 
\hline
$f_\NL= -\sfrac{1}{2}$ & 1.218 & 0.515 & 2.789 & 1.148 & 0.664 & 0.00 & 0.002 \\ 
$f_\NL=-1$ & 1.125 & 0.521 & 2.871 & 1.133 & 0.736& 0.076 & 0.010 \\  
$f_\NL= - \sfrac{3}{2}$ & 0.952 & 0.529 & 2.970 & 1.116 & 0.829 & 0.218 & 0.025 \\
$f_\NL=-2$ & 0.626 & 0.542 & 3.084 & 1.097 & 0.949 & 0.486 & 0.050 \\
\hline
\end{tabular}
\end{large}
\vskip 0.5cm
\end{center}
\end{table}

\begin{figure*} [t!]
\vspace{-1.5cm}
\includegraphics[width=0.49\textwidth]{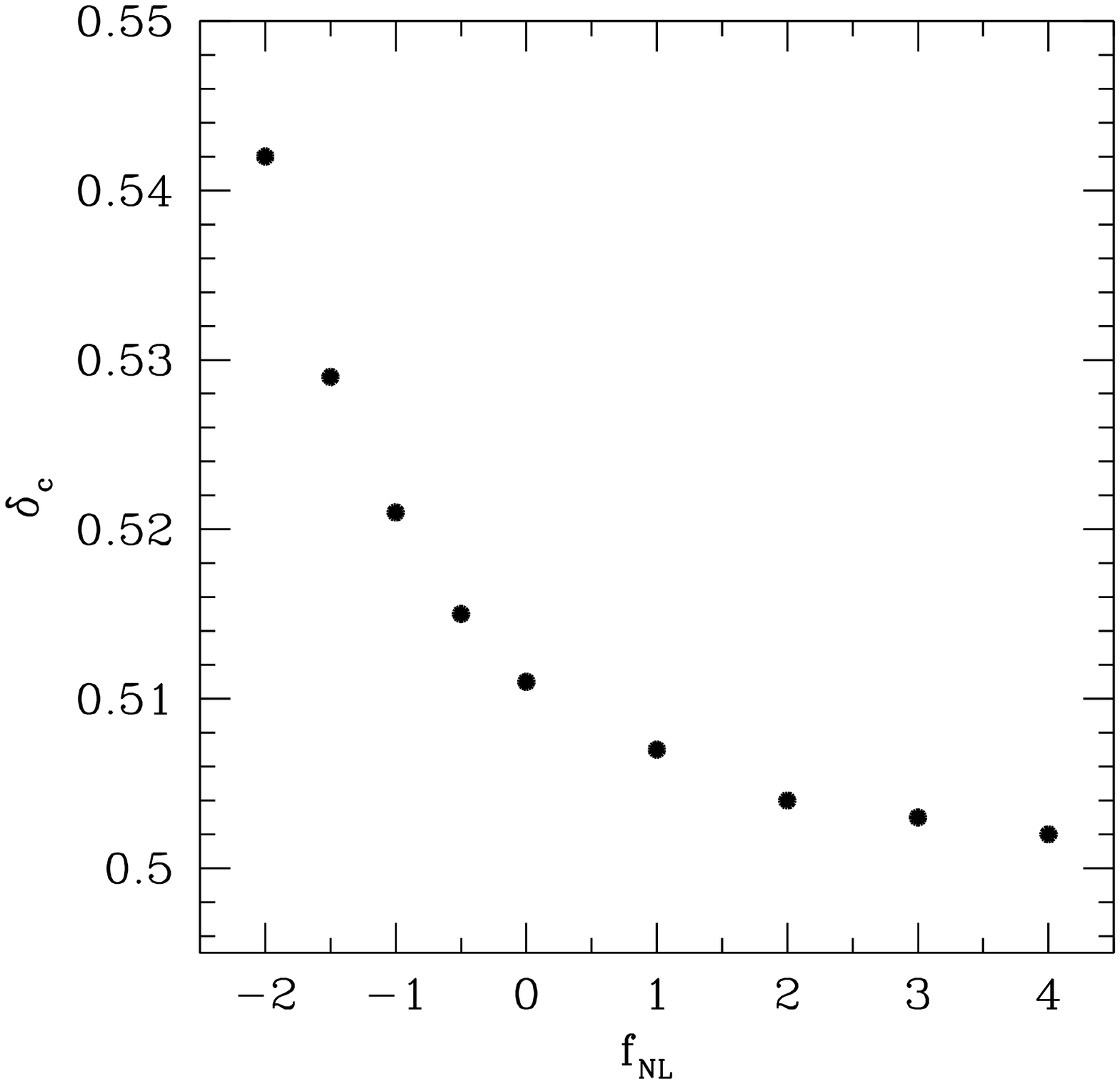} 
  \includegraphics[width=0.49\textwidth]{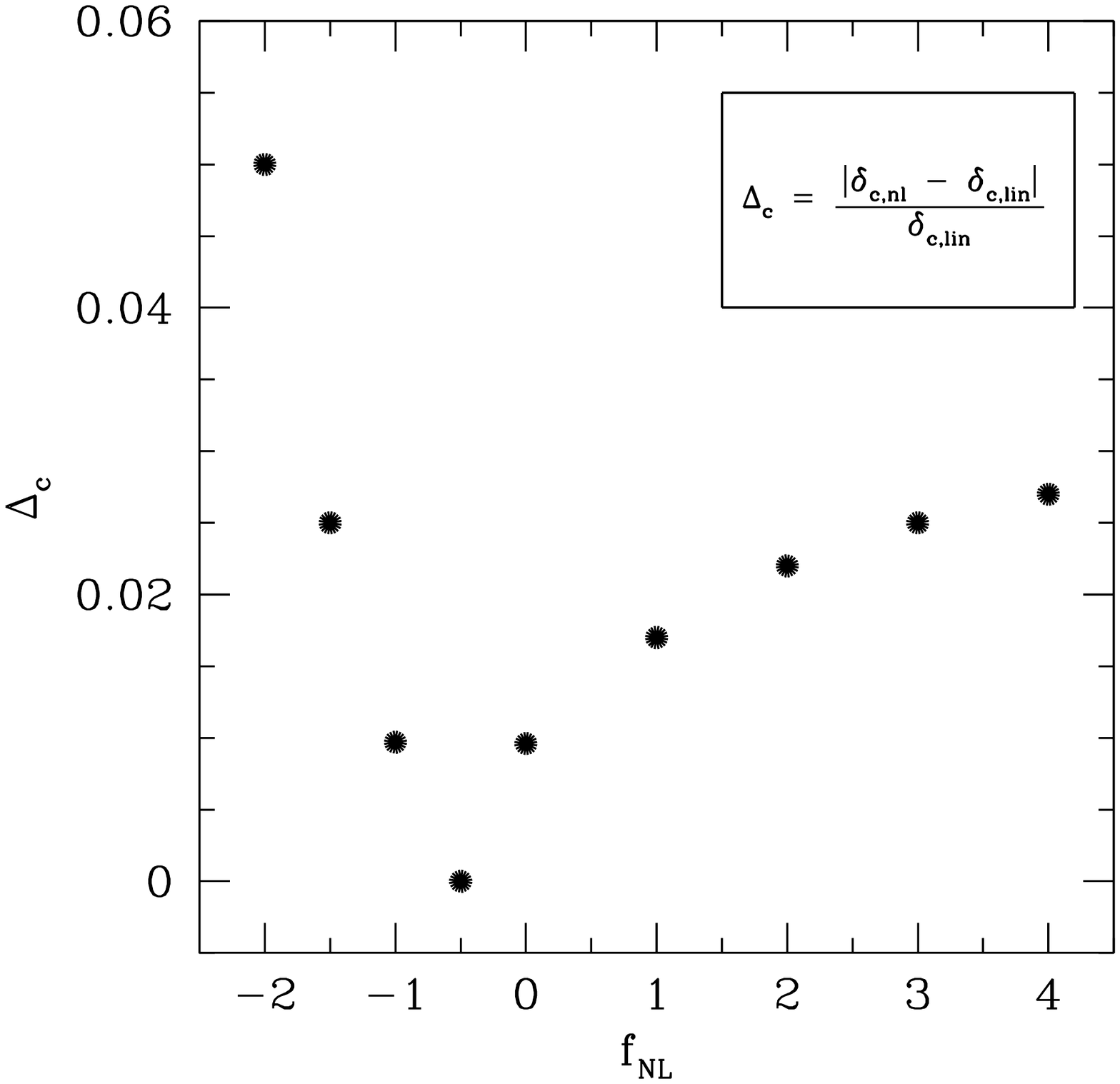} 
   \vspace{-3.0cm}
  \caption{The left plot is showing how threshold $\delta_c$, computed with the linear correction, is varying against $f_\NL$. The right plot is showing the corresponding 
  relative change of $\delta_c$ with respect the value obtained using the linear approximation.}
  \label{delta_c}
\end{figure*}

In the left plot of Figure \ref{delta_c} we are plotting the values of the threshold $\delta_c$ obtained with the non linear corrections, as function of $f_\NL$: $\delta_c$ is 
monotonically decreasing for increasing values of $f_\NL$, converging to $\delta_c\simeq0.5$ for large positive values of $f_\NL$. The corresponding amplitude of the 
critical peak is increasing and converging to a maximum value $\delta\rho_c\simeq1.35$, about $10\%$ larger than the amplitude of the critical peak obtained with the 
linear approximation. On the contrary, for negative values of $f_\NL$, the threshold $\delta_c$ is increasing for $f_\NL$ becoming more and more negative, while the 
amplitude of the critical peak is decreasing. This inverse behaviour of the critical peak amplitude increasing against  the corresponding value of the threshold is due to 
the different amplitude of the pressure gradients modifying the shape during the collapse \cite{musco}. In particular, for $f_\NL=-2$, we find that the density contrast is 
not anymore centrally peaked, and an off-centered peak arises. Looking at Table \ref{table} we see that for negative values of $f_\NL$,  the critical 
amplitude of the density contrast is varying more significantly with respect to the variation obtained for positive values. 

Looking at the right plot of Figure \ref{delta_c}, where we plot the relative change of $\delta_c$ as function of $f_\NL$, we see that this is much more under control 
compared to the critical amplitude of the peak: for the positive values we have analyzed, the variation is less than $3\%$ and the convergent behaviour suggests that it 
will not increase significantly more than this limit, while for negative values of $f_\NL$ the relative change becomes more and more significant, tending to diverge. 
For values of $f_\NL$ down to $-3/2$ the variation is still of few percent, which is consistent with our perturbative approach, but for $f_\NL=2$, when the peak becomes 
off-centered the variation is more significant (about of $5\%$) which is in the limit of what could be considered consistent with a perturbative approach. We argue therefore 
that it is reasonable that the inclusion of higher-order terms in the calculation of the average shape will not change significantly our results for $f_\NL\geq3/2$, while for more 
negative values a significant change could be possible. 

Before concluding, we would like to comment about the approximation done in Section \ref{Averaged density} where we have neglected the exponential term in equation 
\eqref{rho_0}. Using $\delta\rho_{0_G} \sim 1$, as one can see from table \ref{table}, one obtains
\be 
\left( \frac{3}{4} \frac{\delta\rho_{0_G}}{x^2_{m_G}} \right)^3 \frac{{\cal F}(0)}{\pi A} \sim 10^{-3} \left( 1 + \frac{3}{5} f_\NL \right) A^{-1} \,.
\ee
This shows that for values of $A$ large enough, such that the abundance of PBHs is able to explain a significant amount of dark matter, the impact of the exponential term in 
\eqref{rho_0} is quite small and in first approximation can be neglected as we have done for simplicity. However, as we have already explained in the previous section, even 
for values of $A\ll1$ such that this term should be taken into account, the correction of the value of $\delta\rho_{0_G}$ with respect to the peak amplitude $\delta\rho$, is affecting 
significantly only the relative coefficients of the profiles ${\rm Sync}(x), {\cal F}_1(x), {\cal F}_2(x)$, and not the final profile which is a linear combination of similar shapes seen in
Figure \ref{sync}.

\section{Conclusions}
\label{Conslusions}
\noindent
In this paper we have considered the change in the critical threshold induced by the ineludible and intrinsic non-Gaussianity originated by the non-linear relation between the 
overdensity and the curvature perturbation in the case in which the latter is Gaussian. We have also extended our results by assuming a non-Gaussian curvature perturbation. 

The impact of non-Gaussianity in the density contrast threshold, even if the curvature perturbation is Gaussian, alters the PBH abundance. Denoting the change in the threshold  
by $\Delta\delta_c$ and using the expression (\ref{form}),  we can estimate the contribution to the non-linear  abundance from the shift in the threshold to be 
\be
\beta_\NG\simeq e^{-(\delta_c+\Delta\delta_c)^2
/2\sigma^2} \simeq \beta_\G\,e^{-\delta_c\Delta\delta_c
/\sigma^2}=\beta_\G\,e^{-(\delta^2_c
/2\sigma^2)\cdot 2(\Delta\delta_c/\delta_c)}.
\ee
Interesting abundances of PBHs are obtained for $\delta_c/\sigma={\cal O}(6\div 8)$. On the other hand, our results show that the relative change $\Delta\delta_c/\delta_c$ is 
at the percent level (see Figure \ref{delta_c}), leading to a change in the abundance, with respect to the result assuming the Gaussian critical threshold, which is 
$\lesssim{\cal O}(10^2)$. In any case it is smaller than other uncertainties present in the estimate (e.g. the use of statistics), and therefore is not cosmologically significant. 

Our results are based on a perturbative approach which is restricted  to the second-order. Even though limited, we argue that our results should  be rather robust against the 
addition of  higher-order terms for $f_\NL\gsim-3/2$, because the critical threshold is only sensitive to the final shape of the profile of the overdensity, that determines the value
of  the threshold $\delta_c$, and not to the amplitude of the different non linear components. This will be true if the higher order of non-Gaussianity will behave as the $f_\NL$-term, 
not altering significantly the final shape.

We have also seen that the relative change of the threshold $\delta_c$ is much more robust than the relative change of the critical amplitude of the peak $\delta\rho_c$, 
changing up to $10\%$ for positive values of $f_\NL$, and more than $20\%$ for negative values, because the critical amplitude of the peak is much more sensitive to the 
local features of the shape than the threshold $\delta_c$ which is an averaged quantity. We expect that going beyond the perturbative approach, at least for $f_\NL\geq-3/2$, 
will not alter significantly the threshold. This suggest also that the threshold $\delta_c$ would allow to compute the abundance of PBHs with less uncertainties than using the 
local critical amplitude of the peak, as was also pointed out in \cite{Young:2019osy}. 

It will be interesting in the future to understand to which extent the conclusions we have reached here are valid also for a more general shape of the peak of the cosmological 
power spectrum.


\section*{Acknowledgments}
\noindent
 We thank  V. Atal, N. Bellomo, Chris Byrnes, V. De Luca, G. Franciolini, J. Garriga, C. Germani, J. Miller, \mbox{L. Verde} and S. Young for useful discussions. 
A.R. is  supported by the Swiss National Science Foundation (SNSF), project {\sl The Non-Gaussian Universe and Cosmological Symmetries}, project number: 200020-178787.
I.M. is supported by the Unidad de Excelencia Mar\'ia de Maeztu Grant No. MDM-2014-0369. 
 A.K. is supported by the GSRT under EDEIL/67108600.   

\end{document}